\documentclass[fleqn,usenatbib]{mnras}

\usepackage{newtxtext,newtxmath}

\usepackage[T1]{fontenc}

\DeclareRobustCommand{\VAN}[3]{#2}
\let\VANthebibliography\thebibliography
\def\thebibliography{\DeclareRobustCommand{\VAN}[3]{##3}\VANthebibliography}

\usepackage{graphicx}	
\usepackage{amsmath}	

\usepackage{amssymb}	

\title[Concentration of asymmetry in PSBs at $z\sim0.8$]{Central Concentration of Asymmetric Features in Post-starburst Galaxies at $z\sim0.8$}

\author[K. G. Himoto \& M. Kajisawa]{
Kazuharu G. Himoto,$^{1}$
and Masaru Kajisawa,$^{1,2}$\thanks{E-mail: kajisawa@cosmos.phys.sci.ehime.ac.jp}
\\
$^{1}$Graduate School of Science and Engineering, Ehime University, Bunkyo-cho, Matsuyama 790-8577, Japan\\
$^{2}$Research Center for Space and Cosmic Evolution, Ehime University
}

\date{Accepted 2022 December 12. Received 2022 December 11; in original form 2022 September 19}

\pubyear{2022}

\begin{document}
\label{firstpage}
\pagerange{\pageref{firstpage}--\pageref{lastpage}}
\maketitle

\begin{abstract}
We present morphological analyses of Post-starburst galaxies (PSBs)
at $0.7<z<0.9$
in the COSMOS field. We fitted ultraviolet to mid-infrared
multi-band photometry of objects 
with $i<24$ from COSMOS2020 catalogue
with population synthesis models assuming non-parametric, piece-wise 
constant function of star formation history,  
and selected 94 those galaxies that have high specific star
formation rates (SSFRs) of more than $10^{-9.5}$ yr$^{-1}$
in 321--1000 Myr before observation and an order of magnitude lower SSFRs
within recent 321 Myr.
We devised a new non-parametric morphological index
which quantifies concentration of
asymmetric features, $C_{A}$, and measured it as well as
concentration $C$ and asymmetry $A$  
on the {\it Hubble Space Telescope }/Advanced Camera for Surveys
$I_{\rm F814W}$-band images.
While relatively high $C$ and low $A$ values of PSBs are similar with 
those of quiescent galaxies rather than star-forming galaxies, 
we found that PSBs show systematically higher values 
of $C_{A}$ than both quiescent and star-forming galaxies;
36\% of PSBs have $\log{C_{A}} > 0.8$, while only 16\% (2\%) of quiescent
  (star-forming) galaxies show such high $C_{A}$ values. 
Those PSBs with high $C_{A}$ have relatively low overall asymmetry
of $A \sim 0.1$,  
but show remarkable asymmetric features near the centre.
The fraction of those PSBs with high $C_{A}$ increases with increasing
SSFR in 321--1000 Myr before observation rather than residual on-going star
formation. 
These results and their high surface stellar mass densities 
suggest that those galaxies experienced a 
nuclear starburst in the recent past, and processes that cause
such starbursts could lead to 
the quenching of star formation through rapid gas consumption,
supernova/AGN feedback, and so on.
\end{abstract}

\begin{keywords}
galaxies: evolution -- galaxies: formation -- galaxies: structure
\end{keywords}



\section{Introduction}
In general, galaxies can be divided into two populations, namely,
star-forming galaxies (hereafter SFGs) and quiescent galaxies with little
star formation (hereafter QGs).
At $z\lesssim 1$, these two populations show different morphological
properties; QGs tend to show centrally concentrated spheroidal shapes
with little disturbed feature, 
while many SFGs have a (main) disk with spiral patterns and a spheroidal
bulge (e.g., \citealp{rob94}; \citealp{blu19}).
The evolution of number or stellar mass density of these two populations
over cosmic time suggests that some fraction of SFGs stop their
star formation by some mechanism(s) and then evolved into QGs 
(e.g., \citealp{fab07}; \citealp{pen10}).
Such transition from SFG to QG is considered to be one of the most
important processes in galaxy evolution.
While many mechanisms for the quenching of star formation with various decaying 
timescales have been proposed so far (e.g., \citealp{dek86}; \citealp{bar96}; 
\citealp{aba99}; \citealp{bir03}; \citealp{mar09}; \citealp{fab12};
\citealp{spi22}),  
it is still unclear which mechanism plays a dominant role
in galaxy evolution and how it depends on 
conditions such as galaxy properties, environment, epoch, and so on.

Investigating properties of galaxies in the transition phase  
is one of the powerful ways to 
reveal the physical mechanisms of quenching. 
In this context, post starburst galaxies (hereafter, PSBs) that experienced 
a strong starburst followed by quenching in the recent past
have been considered to be an  
important population and studied intensively (see \citealp{fre21}, for recent
 review).
PSBs are selected by their strong Balmer absorption lines and no or weak 
nebular emission lines such as H$\alpha$,[O{\footnotesize II}], and so on
(e.g., \citealp{zab96}; \citealp{dre99}; \citealp{qui04}).
The strong Balmer absorption lines are caused by a significant contribution 
from A-type stars, which indicates high star formation activities  
in the recent past, while no significant emission lines suggest   
little on-going star formation in the galaxy.
While fraction of PSBs is relatively small 
in the entire galaxy population  
($\lesssim $1--2\%, 
\citealp{zab96}; \citealp{qui04}; \citealp{bla04}; \citealp{tra04};
\citealp{got07}; \citealp{wild09};
\citealp{yan09}; \citealp{ver10};
\citealp{won12}; \citealp{row18}; \citealp{che19}), 
PSBs are expected to abruptly stop their star 
formation after starburst and therefore considered to be in a 
 rapid transition phase from SFG to QG. 
Several studies suggest that the fraction of PSBs increases with increasing 
redshift, and a significant fraction of galaxies at $z\sim 1$ 
could pass through the PSB phase when they evolved into quiescent 
galaxies (\citealp{wild09}; \citealp{whi12};
\citealp{wil16}; \citealp{wil20}).

Morphological properties of PSBs have also been investigated so far,  
because they can provide important clues to reveal physical origins 
of the starburst and rapid quenching of star formation. 
Many previous studies found that a significant fraction of PSBs at low and
intermediate redshifts show asymmetric/disturbed features such as
tidal tails (e.g., \citealp{zab96}; \citealp{bla04}; \citealp{tra04};
\citealp{yam05}; \citealp{yan08}; \citealp{pra09}; \citealp{won12};
\citealp{deu20}; \citealp{wil22}).
The morphological disturbances in many PSBs suggest
that galaxy mergers/interactions could be closely related with the
origin of the PSB phase.
Theoretical studies with numerical simulations predicted that 
gas-rich major mergers cause morphological disturbances and gas infall
to the centre of remnants, and then a strong starburst occurs in the central
region (e.g., \citealp{bar91}; \citealp{bar96}; \citealp{bek01}).
Such nuclear starbursts could lead to rapid
quenching of star formation through rapid gas consumption by the burst
and/or gas loss/heating by supernova feedback, AGN outflow, 
tidal force, and so on
(e.g., \citealp{bek05}; \citealp{sny11}; \citealp{dav19};
\citealp{spi22}).

On the other hand, relatively massive PSBs with 
$M_{\rm star} \gtrsim 10^{10} M_{\odot}$ at $z\lesssim 1$
tend to have centrally concentrated early-type morphologies
(e.g., \citealp{qui04}; \citealp{tra04}; \citealp{yam05};
\citealp{yan08}: \citealp{ver10}; \citealp{pra09}; \citealp{mal18}).
These results suggest that morphological changes from those
with a star-forming disk to spheroidal shapes could rapidly proceed,  
if such changes are associated with the transition
through the PSB phase. 
Several studies investigated asymmetric/disturbed features in 
PSBs as a function of time elapsed after starburst, and found
that the disturbed features such as tidal tails weaken or
disappear on a relatively short timescale of $\sim 0.3$ Gyr
(e.g., \citealp{paw16}; \citealp{saz21}).
The numerical simulations of gas-rich mergers also predict
the similar decay of disturbed features on a timescale of
$\sim $ 0.1--0.5 Gyr (\citealp{lot08}; \citealp{lot10};
\citealp{sny15}; \citealp{paw18}; \citealp{nev19}).
While the asymmetric features over entire galaxies seem to
rapidly weaken after starburst, 
observations of CO lines and dust far-infrared emission for
PSBs at low redshifts suggest 
that significant molecular gas and dust are sustained even in
those with relatively old ages of $\sim $ 500-600 Myr
after starburst 
(\citealp{row15}; \citealp{fre18}; \citealp{li19}).
The molecular gas and dust tend to be concentrated in the central
region of those galaxies (e.g., \citealp{sme18}; \citealp{sme22}), 
 and dusty disturbed features near the centre 
in optical images of PSBs have also been reported
(e.g., \citealp{yam05}; \citealp{yan08}; \citealp{pra09}).
Thus such disturbed/asymmetric features in the central region of
PSBs could continue for longer time than those in
outer regions such as tidal tails.

In this paper, we select PSBs that experienced 
a high star formation activity several hundreds Myr before observation 
followed by rapid quenching by performing SED fitting
with multi-band ultraviolet (UV) to mid-infrared (MIR) photometric data
including optical intermediate-bands 
data set in the COSMOS field \citep{sco07a} 
in order to statistically study  morphological properties
of PSBs with relatively 
old ages after starburst at $z \sim 0.8$. 
We investigate their morphological properties 
with non-parametric indices
such as concentration and asymmetry on the
{\it Hubble Space Telescope }/Advanced Camera for Surveys ({\it HST}/ACS) data,
in particular, focusing on concentration of
asymmetric features near the centre of these galaxies.
Section 2 describes the data used in this study. In Section 3,
we describe sample selection with the SED fitting and
 methods to measure the non-parametric morphological indices,
including newly devised concentration of asymmetric features.
In Section 4, we present the morphological indices of PSBs 
and compare them with those of SFGs and QGs. We discuss 
our results and their implications in Section 5, and summarise the
results of this study in Section 6.
Throughout this paper, 
 we assume a flat universe with $\Omega_{\rm matter}=0.3$, $\Omega_{\Lambda}=0.7$,
and $H_{0}=70$ km s$^{-1}$ Mpc$^{-1}$, and magnitudes are given in the AB system.

\section{Data} \label{sec:data}

In this study, we used a multi-wavelength catalogue of COSMOS2020 
\citep{wea22} to construct a sample of $z\sim0.8$ galaxies that
experienced a starburst followed by rapid quenching several hundreds Myr
before observation. 
\citet{wea22} provided multi-band photometry from UV to MIR 
wavelengths, namely, $GALEX$ FUV and NUV \citep{zam07}, CFHT/MegaCam $u$ and $u^{*}$ \citep{saw19}, Subaru/HSC $grizy$ \citep{aih19}, Subaru/Suprime-Cam $Bg^{'}Vr^{'}i^{'}z^{'}z^{''}$ \citep{tan07} and 12 intermediate and 2 narrow bands
\citep{tan15}, VISTA/VIRCAM $YJHK_{s}$ and $NB119$ \citep{mcc12}, and
$Spitzer$/IRAC ch1--4 (\citealp{ash13}; \citealp{ste14}; \citealp{ash15}: \citealp{ash18}), for objects in the COSMOS field \citep{sco07a}. 
Source detection was performed on an $izYJHK_{s}$-band combined image, and
aperture photometry was done on PSF-matched images with SExtractor
\citep{ber96}.
We used the photometry measured with 3 arcsec diameter apertures   
in the $GALEX$ NUV, MegaCam $u$
and $u^{*}$, HSC $grizy$, Suprime-Cam $BVi^{'}z^{''}$ and 12 intermediate, 
VIRCAM $YJHK_{s}$, and IRAC ch1--4 bands for objects with $i<24$ from 
``CLASSIC'' catalogue of COSMOS2020 in order 
to carry out SED fitting analysis and sample selection
described in Section \ref{sec:ana}.
We set the magnitude limit of $i<24$ 
to investigate their detailed morphology
on {\it HST}/ACS data and
ensure accuracy of photometry, in particular,
that of the intermediate bands in the SED fitting.
We excluded X-ray AGNs from our sample because the SED fitting described
in the next section does not include AGN emission.
The photometric offsets derived from the SED fitting with LePhare 
\citep{wea22} were applied to these fluxes. 
Furthermore, we similarly estimated additional small photometric offsets
by performing 
the SED fitting described in the next section for isolated galaxies
with spectroscopic redshifts from zCOSMOS (\citealp{lil07}; \citealp{lil09})
and LEGA-C (\citealp{van16}; \citealp{van21}) surveys,
and applied them. 
We corrected these fluxes for Galactic extinction by using $E(B-V)$ value
of the Milky Way at each object position from the catalogue.

For morphological analysis,  
we used COSMOS {\it HST}/ACS $I_{\rm F814W}$-band data version 2.0 \citep{koe07}. 
The data have a pixel scale of 0.03 arcsec/pixel and a PSF FWHM  
of $\sim 0.1$ arcsec.
The 50\% completeness limit is $I\sim 26$ mag for sources with a
half-light radius of 0.5 arcsec \citep{sco07b}. 
We also used the Subaru/Suprime-Cam $i^{'}$-band
data in order to determine pixels of the ACS image which belong to the
target galaxy. The reduced $i^{'}$-band data have a pixel scale of
0.15 arcsec and a PSF FWHM of $\sim 1.0$ arcsec \citep{tan07}.

\section{Analysis} \label{sec:ana}

\subsection{SED fitting} \label{sec:sed}

\begin{figure} 
  \includegraphics[width=\columnwidth]{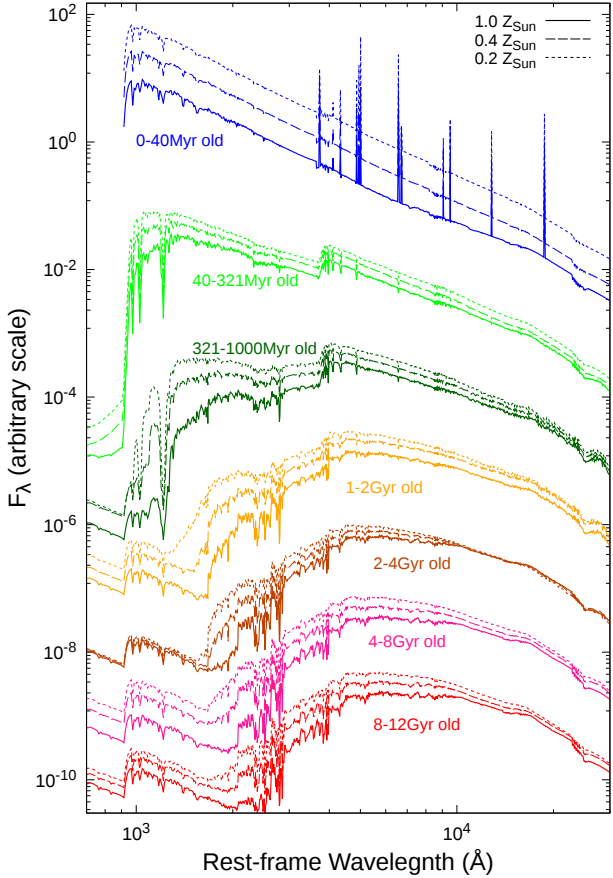}
  \caption{The model templates of stars formed in the seven periods of
     look-back time. They are constructed from the GALAXEV library under
    the assumption of constant SFR in each period.
    The linear combination of these templates is fitted 
    to the observed SEDs. The short-dashed, long-dashed, and solid lines
    show those with 0.2, 0.4, and 1.0 $Z_{\odot}$, respectively.
    \label{fig:temp}}
\end{figure}

In order to select PSBs that experienced a high star formation activity
followed by rapid quenching several hundreds Myr before observation,
we fitted the multi-band photometry of objects with $i<24$ from the
COSMOS2020 catalogue mentioned above with population 
synthesis models of GALAXEV \citep{bru03}.
For the purpose, we adopted non-parametric, piece-wise constant function of
star formation history (SFH) where SFR varies among different time intervals
but is constant in each interval,  
following previous studies (\citealp{toj07}; \citealp{kel14};
\citealp{lej17}; \citealp{lej19}; \citealp{cha18}).
We divided the look-back time for each galaxy into seven periods, namely,
0--40 Myr, 40--321 Myr, 321--1000 Myr, 1--2 Gyr, 2--4 Gyr, 4--8 Gyr, and
8--12 Gyr before observation, which are similar with those used in
\citet{lej17}. We constructed model SED templates of stars formed in
the different periods by assuming Chabrier IMF
\citep{cha03} and constant SFR in each period.
Figure \ref{fig:temp} shows the model templates of stars formed in the
seven periods for different metallicities.
The model SEDs used in the fitting are based on a linear combination of
the seven templates, and normalisation coefficients for the seven 
templates are free parameters. Thus SFHs are expressed as constant SFRs in
the seven time intervals.

In order to search the best-fit values of the normalisation
coefficients that provide the minimum $\chi^{2}$, 
we adopted Non-Negative Least Squares (NNLS) algorithm
\citep{law74} following GASPEX by \citet{mag15}, 
while we used a simple full grid search for redshift, metallicity, and dust
extinction.
The NNLS algorithm basically solves a system of linear equations represented
as a matrix equation, but carries out some iterations to search non-negative
solutions with changing non-active (zero-value) coefficients.
The templates with three stellar metallicities, namely, 0.2, 0.4, and 1.0
$Z_{\odot}$ were fitted.
If we include templates with 2.5 $Z_{\odot}$, our sample of PSBs is
almost the same and results in this study do not change. 
For simplicity, we fixed the metallicity over all the periods
except for the youngest one, 0--40 Myr before observation.
The metallicity of the template of 0--40 Myr was independently chosen
from the same 0.2, 0.4, and 1.0 $Z_{\odot}$.
We added nebular emission only in the youngest template by using PANHIT
(\citealp{maw16}; \citealp{maw20}) because a contribution from
the nebular emission is negligible in templates of the other
older periods.
In PANHIT, the nebular emission is calculated from ionising spectra of
the stellar templates following \citet{ino11}.
We used only nebular continuum and hydrogen recombination lines 
from PANHIT,
and recalculated fluxes of the other emission lines, namely
[O{\footnotesize II}]$\lambda\lambda$3727, [O{\footnotesize III}]$\lambda\lambda$4959,5007, [N{\footnotesize II}]$\lambda\lambda$6548,6583,
[S{\footnotesize II}]$\lambda\lambda$6726, and
[S{\footnotesize III}]$\lambda\lambda$9069,9532 by using emission
  line ratios of local star-forming galaxies with various gas metallicities
  (\citealp{nag06}; \citealp{vil96}), because relatively high 
  [O{\footnotesize III}]/[O{\footnotesize II}] ratios 
  in \citet{ino11} model at high metallicity lead to
 slight underestimates of 
 photometric redshift for some fraction of star-forming galaxies
at $z\sim$ 0.5--1.0.
This is because the intermediate bands of Subaru/Suprime-Cam densely
sample wavelengths around the Balmer break at these redshifts, and 
 the deficit of the model [O{\footnotesize II}] fluxes 
in an intermediate band mimics the Balmer break at slightly lower redshifts. 
In the calculation of these emission lines, 
we assumed the same gas metallicity with the stellar one
(i.e., 0.2, 0.4, and 1.0 $Z_{\odot}$) as in PANHIT.
The fraction of ionising photons that really ionise gas
rather than escape from the galaxy or are absorbed by dust 
is also varied from 0.1 to 1.0, while the fraction does not
affect results in this study. 

Linear independence among the templates is important for 
estimating SFHs of galaxies accurately (e.g., \citealp{mag15}). 
In order to select galaxies that experienced a high star formation
activity followed by
rapid quenching within $\sim 1$ Gyr, we chose the intervals of
the periods younger than 1 Gyr so that these templates are
close to be orthogonal for wavelength resolutions of the intermediate
bands ($\lambda/\Delta\lambda \sim 20$).
In Figure \ref{fig:temp}, one can see that 
variations in SED among different periods are relatively large
for those with young ages of $< 1$ Gyr, while  
differences in metallicity do not so strongly affect SEDs.
In appendix, we calculated inner products between 
the templates with different periods/metallicities to examine the linear independence,
and confirmed that 'angles' between the templates with different periods
($<$ 1 Gyr) are relatively large, while those of the same period
with different metallicities are near-parallel. 
Thus we expect that SFH at $<$ 1 Gyr can be reproduced relatively well,
while our assumption of the fixed metallicity except for the youngest
period could affect our selection for PSBs if the metallicity significantly
changed between different periods. 
On the other hand, the variations in SED among the different
periods and metallicities are much smaller for older ages of $>$ 
1 Gyr, which is well known as age-metallicity degeneracy \citep{wor94}.
By setting the several periods at $> 1$ Gyr,
we intend to keep flexibility to avoid systematic
effects of a single long period of the old age on the fitting in the
younger periods, while we do not aim to accurately estimate details of 
SFH and metallicity at $> 1$ Gyr.

For the dust extinction, we used the Calzetti law \citep{cal00} 
and attenuation curves for local star-forming galaxies with different stellar
masses, namely $10^{8.5}$--$10^{9.5} M_{\odot}$, $10^{9.5}$--$10^{10.5} M_{\odot}$,
and $10^{10.5}$--$10^{11.5} M_{\odot}$, from \citet{sal18}.
These four attenuation curves allow us to 
 cover observed variation in 2175\AA\ bump and reproduce  
 correlation between the bump strength and overall slope
of the attenuation curve \citep{sal18}.
We adopted different ranges of $E(B-V)$ (or $A_{V}$) for these different
attenuation curves, namely, $E(B-V) \leq 1.6$ for the Calzetti law and
$E(B-V) \leq 0.4$ for those from \citet{sal18}, 
to take account of observed correlation between
the overall slope and $V$-band attenuation, i.e., the slope tends to be  
steeper at smaller optical depth (\citealp{sal18}; \citealp{sal20}).

The intergalactic matter (IGM) absorption of 
\citet{mad95} was applied to the dust-reddened model SEDs at each redshift. 
We took variation of the IGM absorption among lines of sight
into account by adding fractional errors to fluxes in bands
 shorter than the rest-frame 1216 \AA\ as in \citet{yon22}.

In the fitting, we searched the minimum $\chi^{2}$ at each redshift,
and calculated a redshift likelihood function,
$P(z) \propto \exp(-\frac{\chi^{2}(z)}{2})$,
where $\chi^{2}(z)$ is the minimum $\chi^{2}$ at each redshift.
We adopted the median of the likelihood function as a redshift of each
object (e.g., \citealp{ilb10}).
The photometric data with a central rest-frame wavelength longer than 25000 \AA\ 
were excluded from the calculation of the minimum $\chi^{2}$ at each redshift, 
because our model templates do not include the dust/PAH emission.
We checked accuracy of the estimated redshifts by using the spectroscopic
redshift catalogues from LEGA-C and zCOSMOS.
At $z_{\rm spec} < 1$, 2587 and 6704 spectroscopic redshifts from LEGA-C and
zCOSMOS were matched to our sample.
The fraction of those with $\Delta z/(1+z_{\rm spec}) > 0.1$ is very small
(1.35\% and 0.45\% for the LEGA-C and zCOSMOS samples, respectively), 
and the means and standard deviations of $\Delta z/(1+z_{\rm spec})$ for
galaxies with $\Delta z/(1+z_{\rm spec}) < 0.1$ are $-0.009 \pm 0.011$
and $-0.008 \pm 0.012$ for the LEGA-C and zCOSMOS samples, respectively.
For PSBs described in the next subsection, 27 spectroscopic redshifts 
from the both catalogues are available at $z_{\rm spec} < 1$, 
and the accuracy of the estimated
redshifts is almost the same as all the spec-$z$ sample with no outlier of 
$\Delta z/(1+z_{\rm spec}) > 0.1$.
Such high photometric redshift accuracy is provided by the optical
intermediate-bands data, which densely sample wavelengths around
the Balmer/4000\AA\ break, and is consistent with those in
previous studies (\citealp{lai16}; \citealp{wea22}).
Including the attenuation curves from \citet{sal18} with different
slopes and UV bump strengths also contributes
to the accuracy of the photometric redshifts.
If we use only the Calzetti attenuation law with a fixed slope and no
UV bump, the fraction of catastrophic failure increases by $\sim 40$ \%
and the standard deviation of $\Delta z/(1+z_{\rm spec})$ increases by
$\sim 30$ \%.

We carried out Monte Carlo simulations to estimate probability
distributions of derived physical properties such as stellar mass at
the observed epoch, SFR and SSFR in the periods of look-back time.
We added random shifts based on photometric errors to the observed
fluxes and then performed the same SED fitting with the simulated photometries
fixing the redshift to the value mentioned above.
We did 1000 such simulations for each object and adopted the median values
and 68\% confidence intervals as the physical properties and
their uncertainty, respectively.
\begin{figure}
  \includegraphics[width=\columnwidth]{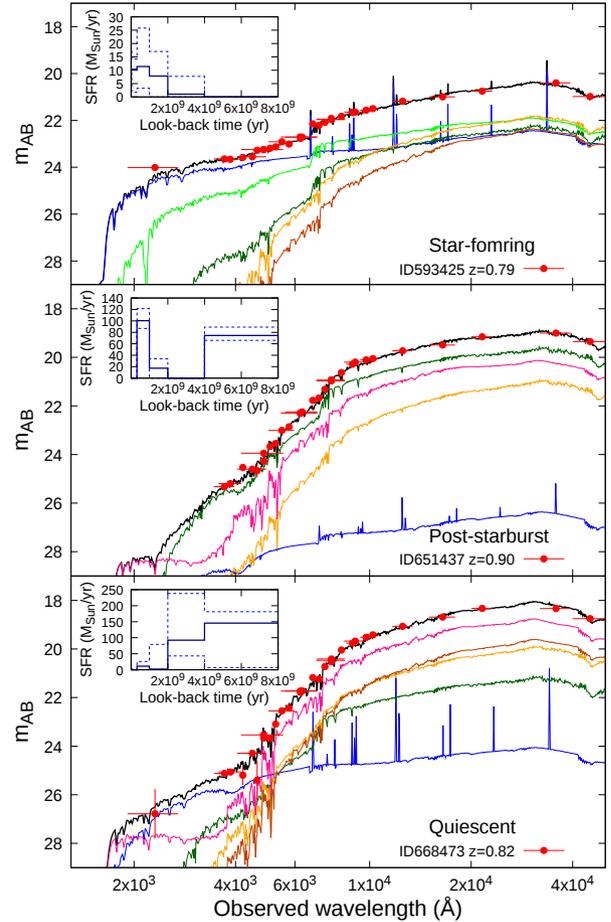}
  \caption{ Examples of the SED fitting for galaxies at $z\sim0.8$.
    Red circles show the observed fluxes, and vertical and horizontal
    error bars represent flux errors and FWHMs of the bands, respectively.
    The thick solid line shows the best-fit model SED, while thin lines 
    represent contributions from stars formed in the different periods.
    The colours of these thin lines are the same as Figure \ref{fig:temp}.
    The estimated SFHs are shown in the insets, where solid and dashed lines
    represent the best-fit SFR and its 68\% confidence interval, respectively.
    These three examples show different types of SFHs, namely,
    SFG, PSB, and QG (see text for selection criteria).
  \label{fig:sedexam}}
\end{figure}
In Figure \ref{fig:sedexam}, we show examples of the SED fitting for
galaxies at $z\sim0.8$ classified into different types of SFHs
in the next section. The contributions from stars formed in
the different periods in 
the best-fit models and estimated SFHs are shown in the figure.

\subsection{Sample selection} \label{sec:sample}

\subsubsection{Post-Starburst Galaxies}

We used the physical properties estimated from
the SED fitting in the previous subsection 
to select galaxies that experienced active star formation followed by rapid
quenching several hundreds Myr before observation. 
We set selection criteria with SSFRs in the three youngest periods of 
look-back time, namely, SSFR$_{\rm 0-40Myr}$, SSFR$_{\rm 40-321Myr}$, and
SSFR$_{\rm 321-1000Myr}$.
Note that we define these SSFRs as SFRs in these periods divided by
stellar mass at the observed epoch (e.g.,
SSFR$_{\rm 0-40Myr} = $ SFR$_{\rm 0-40Myr}/M_{\rm star, 0}$) to easily compare the SSFRs
among the different periods.
Thus the SSFRs used in this study do not coincide with exact values of
SFR$/M_{\rm star}$ in these periods.
Our selection criteria are
\begin{equation}
  \begin{split}
  &{\rm SSFR}_{\rm 321-1000Myr} > 10^{-9.5} \hspace{1mm} {\rm yr}^{-1} \hspace{1mm} \&\\   
  &{\rm SSFR}_{\rm 40-321Myr} < 10^{-10.5} \hspace{1mm} {\rm yr}^{-1} \hspace{1mm} \&\\ 
  &{\rm SSFR}_{\rm 0-40Myr} < 10^{-10.5} \hspace{1mm} {\rm yr}^{-1}.
  \end{split}
  \label{eq:psbsel}
  \end{equation}
This selection utilises a characteristic SED shape of stars formed in   
321--1000 Myr before observation, namely, strong Balmer break,
steep and red FUV and relatively flat NUV continuum, and relatively
blue continuum at longer wavelength than the break (Figure \ref{fig:temp}).
We used those galaxies at $0.7<z<0.9$ in this study, because 
the intermediate bands densely sample the rest-frame NUV to $B$ band
for these redshifts and enable us to distinguish the characteristic
SED of these stars from SEDs of stars formed in the other periods.
Furthermore, the {\it HST}/ACS $I_{\rm F814W}$ band corresponds to  
the rest-frame $B$ band at $z\sim0.8$. Since many morphological studies
of galaxies have been done in the rest-frame $B$ band, we can easily
compare our morphological results with the previous studies.

Since distributions of SSFR in the periods of 0--40 Myr, 40--321 Myr,
and 321--1000 Myr for galaxies with $i<24$ at $0.7<z<0.9$
 show a peak at $\sim 10^{-9.5}$ yr$^{-1}$,
our selection picks up those galaxies whose SSFRs were comparable to or
higher than the main sequence of SFGs in 321--1000 Myr
before observation and then decreased at least by an order of magnitude 
in the last 321 Myr.
We excluded galaxies with the reduced minimum $\chi^{2} > 5$ from our analysis,
because the estimated SSFRs are unreliable for those objects.
We also discarded objects affected by nearby bright sources
in the {\it HST}/ACS
$I_{\rm F814W}$-band images, and confirmed that all remaining galaxies in our
sample are not saturated in the $I_{\rm F814W}$-band images.
Finally, there are 17459 galaxies with $i<24$ and reduced $\chi^{2} <5$
at $0.7<z<0.9$ in the {\it HST}/ACS $I_{\rm F814W}$-band data,
and we selected 94 PSBs. 

\begin{figure}
  \includegraphics[width=\columnwidth]{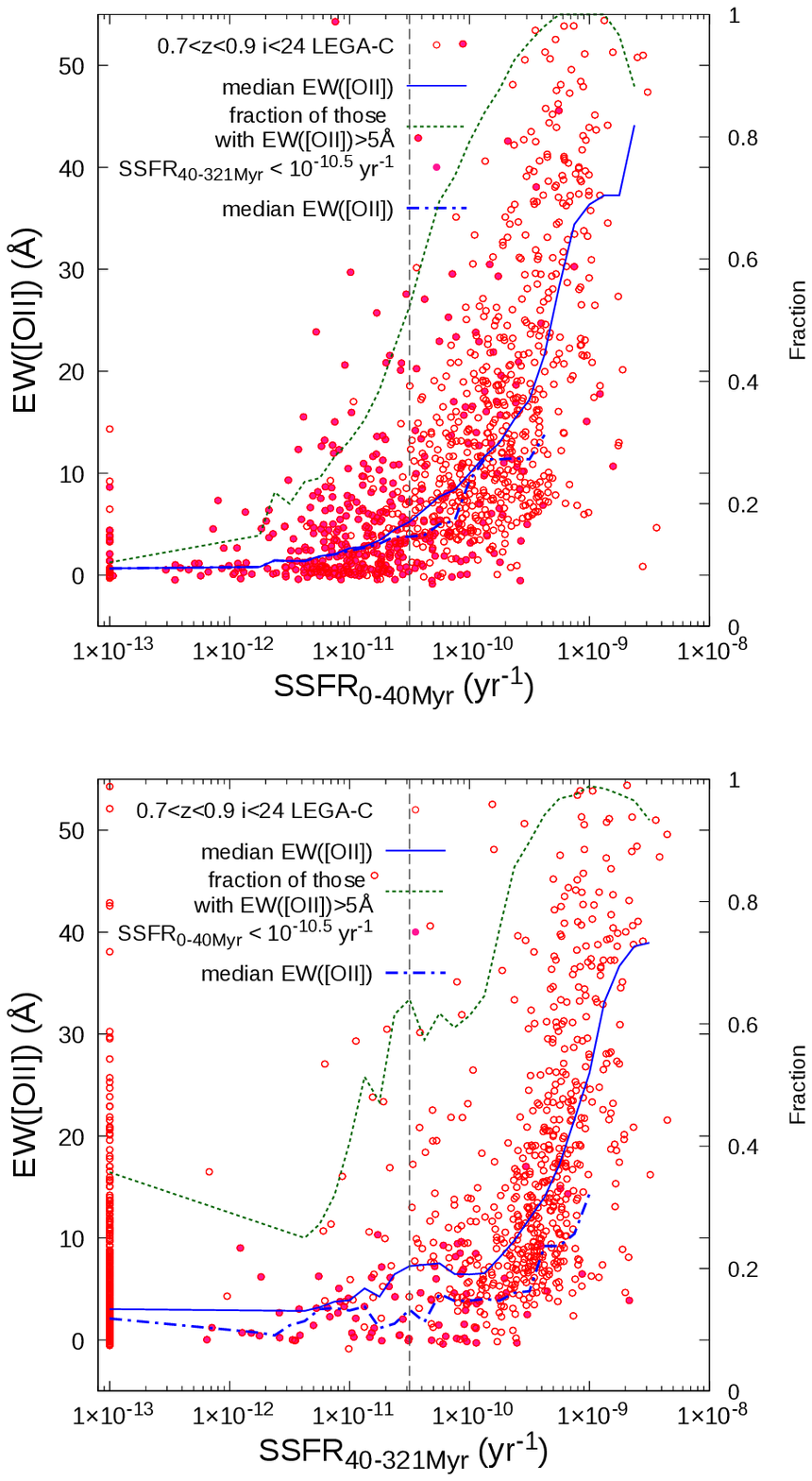}
  \caption{
    EW([OII]) as a function of SSFR$_{\rm 0-40Myr}$ (top) and
    SSFR$_{\rm 40-321Myr}$ (bottom) for our sample galaxies with the spectroscopic
    measurements by LEGA-C \citep{van21}.
    Open circles show all the galaxies with $i<24$ at $0.7<z<0.9$, and solid
   circles represent those with SSFR$_{\rm 40-321Myr} < 10^{-10.5}$ yr$^{-1}$
    in the top panel and those with SSFR$_{\rm 0-40Myr} < 10^{-10.5}$ yr$^{-1}$
    in the bottom panel.
    Those with SSFR $ < 10^{-13}$ yr$^{-1}$ (SSFR $= 0$
    for most cases) are plotted at $10^{-13}$ yr$^{-1}$.
    The solid and dashed-dotted lines show the median
    values of EW([OII]) in SSFR bins with a width of $\pm$ 0.25 dex
    for all the galaxies and those with
    SSFR$_{\rm 40-321Myr} < 10^{-10.5}$ yr$^{-1}$ (top panel) or 
    SSFR$_{\rm 0-40Myr} < 10^{-10.5}$ yr$^{-1}$ (bottom panel), respectively.
    The short-dashed line shows the fraction of those with
    EW([OII]) $> $ 5\AA.
    The vertical long-dashed line shows the boundary of SSFR $ = 10^{-10.5}$
    yr$^{-1}$, and selected PSBs have lower SSFRs than this value.  
  \label{fig:OII}}
\end{figure}

\begin{figure}
  \includegraphics[width=\columnwidth]{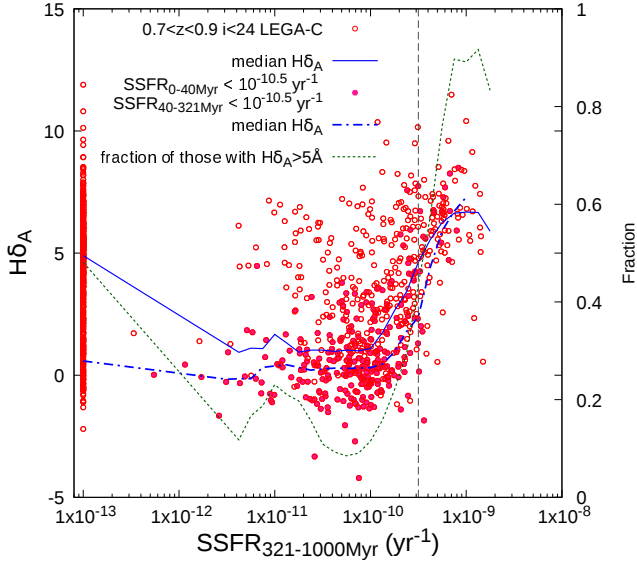}
  \caption{
    H$\delta_{\rm A}$ as a function of SSFR$_{\rm 321-1000Myr}$
    for the sample galaxies
    with the spectroscopic measurements by LEGA-C.
    Open circles show all the galaxies with $i<24$ at $0.7<z<0.9$, and
    solid circles represent those with SSFR$_{\rm 0-40Myr} < 10^{-10.5}$ yr$^{-1}$
    and SSFR$_{\rm 40-321Myr} < 10^{-10.5}$ yr$^{-1}$.
    Those with SSFR$_{\rm 321-1000Myr} < 10^{-13}$ yr$^{-1}$ are plotted at
    $10^{-13}$ yr$^{-1}$.
    The solid and dashed-dotted lines show the median values of H$\delta_{\rm A}$
    in SSFR$_{\rm 321-1000Myr}$ bins with a width of $\pm$ 0.25 dex for
    all the galaxies and those with SSFR$_{\rm 0-40Myr} < 10^{-10.5}$ yr$^{-1}$
    and SSFR$_{\rm 40-321Myr} < 10^{-10.5}$ yr$^{-1}$.
    The short-dashed line shows the fraction of those with
    H$\delta_{\rm A} > $ 5\AA.
    The vertical long-dashed line shows the boundary of
    SSFR$_{\rm 321-1000Myr} = 10^{-9.5}$ yr$^{-1}$, and selected PSBs have higher
    SSFR$_{\rm 321-1000Myr}$ than this value.
  \label{fig:Hd}}
\end{figure}

In order to check relation between our selection method
and the spectroscopic 
selection for PSBs used in previous studies,
we compared the estimated SSFRs in the youngest three periods
with spectroscopic
indices of [O{\footnotesize II}] emission and H$\delta$ absorption lines
from the LEGA-C survey \citep{van21}. 
For the purpose, we searched for galaxies with $i<24$ and the
reduced $\chi^{2} < 5$ at $0.7<z<0.9$
in the LEGA-C catalogue, and found 1265 matched objects.
We used EW([O{\footnotesize II}]) and H$\delta_{\rm A}$ indices from
the LEGA-C catalogue. The H$\delta_{\rm A}$ index is corrected for contribution
from the emission line \citep{van21}.
The upper panel of Figure \ref{fig:OII} shows 
EW([O{\footnotesize II}]) as a function of SSFR$_{\rm 0-40Myr}$.
One can see that EW([O{\footnotesize II}]) tends to increase with
increasing SSFR$_{\rm 0-40Myr}$, while there is a relatively large scatter at
a given SSFR$_{\rm 0-40Myr}$.
At SSFR$_{\rm 0-40Myr} < 10^{-10.5}$ yr$^{-1}$, the median EW([O{\footnotesize II}])
becomes $\lesssim 5$ \AA, which is often used as one of the spectroscopic
criteria for PSBs, and the fraction of those with 
EW([O{\footnotesize II}]) $ > 5$ \AA\ is less than $\sim 0.5$. 
Solid circles in the panel show those with 
SSFR$_{\rm 40-321Myr} < 10^{-10.5}$ yr$^{-1}$, and therefore
those solid circles at SSFR$_{\rm 0-40Myr} < 10^{-10.5}$ yr$^{-1}$ 
represent objects whose both SSFR$_{\rm 0-40Myr}$ and SSFR$_{\rm 40-321Myr}$
are lower than $10^{-10.5}$ yr$^{-1}$. 
Their median EW([O{\footnotesize II}]) is less than $\sim 4$ \AA\ 
at SSFR$_{\rm 0-40Myr} < 10^{-10.5}$ yr$^{-1}$. 
The bottom panel of Figure \ref{fig:OII} shows
relation between SSFR$_{\rm 40-321Myr}$ and EW([O{\footnotesize II}]).
The similar trend as in the upper panel can be seen, although   
the median EW([O{\footnotesize II}]) at
SSFR$_{\rm 40-321Myr} \sim 10^{-10.5}$ yr$^{-1}$ 
is slightly higher ($\sim 7$ \AA).

 The median EW([O{\footnotesize II}]) for those galaxies with both 
 SSFR$_{\rm 0-40Myr} < 10^{-10.5}$ yr$^{-1}$ and SSFR$_{\rm 40-321Myr} < 10^{-10.5}$
 yr$^{-1}$ is only $\sim 1.3$ \AA, and the fraction of those with
 EW([O{\footnotesize II}]) $ > 5$ \AA\ is $\sim 0.24$.
Thus the criteria of low SSFR$_{\rm 0-40Myr}$ and SSFR$_{\rm 40-321Myr}$ can 
select those galaxies with low EW([O{\footnotesize II}]), although
some fraction of galaxies with EW([O{\footnotesize II}]) $ > 5$ \AA\ can
be included in our sample.
The relatively high EW([O{\footnotesize II}]) seen in some of those galaxies
could be caused by shock or weak AGN rather than star formation, 
as some previous studies reported that a fraction of
``H$\delta$ strong'' galaxies with rapidly declining SFHs  
show detectable [O{\footnotesize II}] emission lines 
(e.g., \citealp{yan06}; \citealp{paw18}).

\begin{figure*}
  \includegraphics[width=2.0\columnwidth]{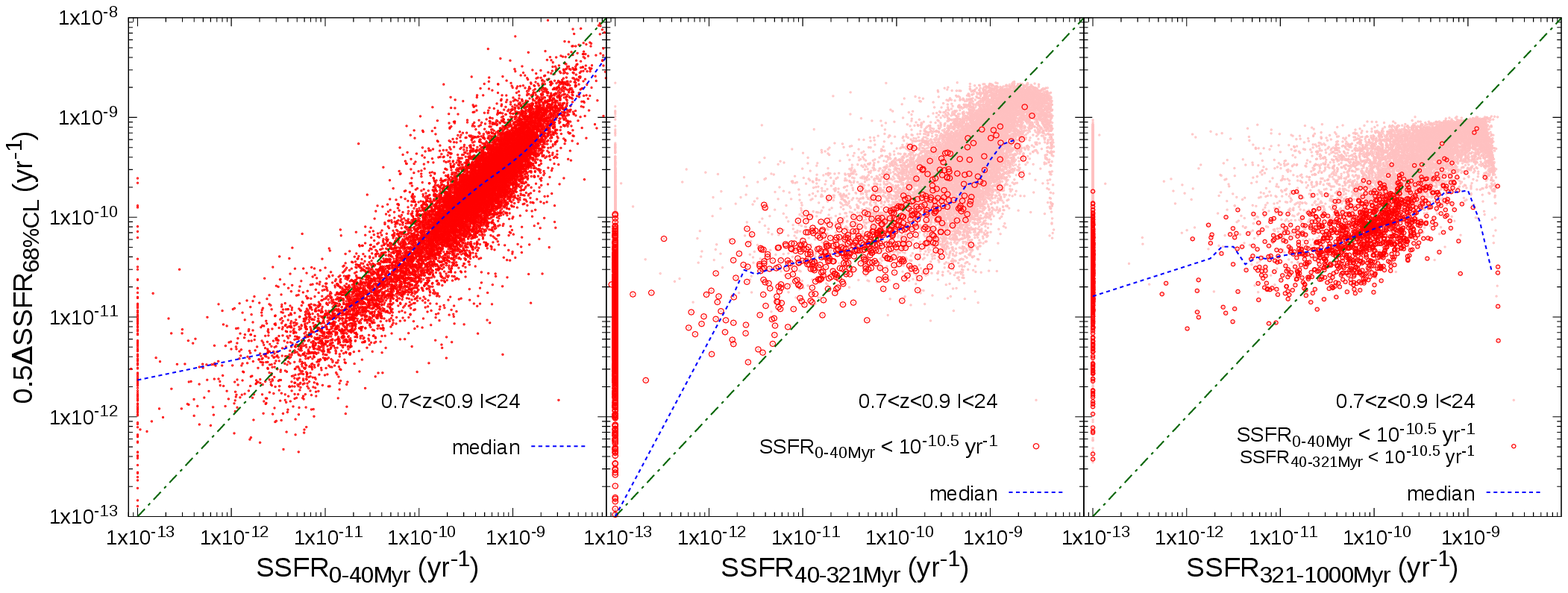}
  \caption{
    {\bf left:} The uncertainty of SSFR$_{\rm 0-40Myr}$ estimated in the SED
    fitting for all the sample galaxies with $i<24$ at $0.7<z<0.9$.
    Half widths of the 68\% confidence interval of SSFR$_{\rm 0-40Myr}$
    are shown as a function of SSFR$_{\rm 0-40Myr}$ itself.
    Those with SSFR$_{\rm 0-40Myr} < 10^{-13}$ yr$^{-1}$ are plotted at
    $10^{-13}$ yr$^{-1}$.
    The dashed line shows the median values in SSFR$_{\rm 0-40Myr}$ bins with
    a width of $\pm$ 0.125 dex.
    {\bf middle:} The same as the left panel but for SSFR$_{\rm 40-321Myr}$.
    While magenta dots show all the galaxies with $i<24$ at $0.7<z<0.9$,
    red open circles represent those with
    SSFR$_{\rm 0-40Myr} < 10^{-10.5}$ yr$^{-1}$. 
    The dashed line shows the median values in SSFR$_{\rm 40-321Myr}$ bins for
    those with SSFR$_{\rm 0-40Myr} < 10^{-10.5}$ yr$^{-1}$.
    {\bf right:} The same as the left panel but for SSFR$_{\rm 321-1000Myr}$.
    Magenta dots show all the sample galaxies, and red open circles represent
    those with SSFR$_{\rm 0-40Myr} < 10^{-10.5}$ yr$^{-1}$ and
    SSFR$_{\rm 40-321Myr} < 10^{-10.5}$ yr$^{-1}$.
    The dashed line shows the median values in SSFR$_{\rm 321-1000Myr}$ bins for
    those with SSFR$_{\rm 0-40Myr} < 10^{-10.5}$ yr$^{-1}$ and
    SSFR$_{\rm 40-321Myr} < 10^{-10.5}$ yr$^{-1}$.
  \label{fig:ssfrerr}}
\end{figure*}

Figure \ref{fig:Hd} shows H$\delta_{\rm A}$ as a function of
SSFR$_{\rm 321-1000Myr}$. 
The H$\delta_{\rm A}$ index increases with increasing SSFR$_{\rm 321-1000Myr}$
at SSFR$_{\rm 321-1000Myr} \gtrsim 10^{-10}$ yr$^{-1}$.
The median H$\delta_{\rm A}$ is $\sim 1$ \AA\ at
SSFR$_{\rm 321-1000Myr}\sim 10^{-10}$ yr$^{-1}$ and increases to $\sim 7$\AA\
around SSFR$_{\rm 321-1000Myr} \sim 10^{-9}$ yr$^{-1}$.
A relatively high median H$\delta_{\rm A}$ value at
SSFR$_{\rm 321-1000Myr} = 10^{-13}$ yr$^{-1}$  
(SSFR$_{\rm 321-1000Myr} = 0$ for most cases) is caused by those galaxies
with high SSFR$_{\rm 0-40Myr}$ and/or SSFR$_{\rm 40-321Myr}$.
When we limit to those galaxies with SSFR$_{\rm 0-40Myr} < 10^{-10.5}$ yr$^{-1}$
and SSFR$_{\rm 40-321Myr} < 10^{-10.5}$ yr$^{-1}$ (solid circles in the figure),  
the median H$\delta_{\rm A}$ is 
$\sim 1$ \AA\ at SSFR$_{\rm 321-1000Myr} \lesssim 10^{-10.5}$ yr$^{-1}$, and
the rapid increase at SSFR$_{\rm 321-1000Myr} > 10^{-10}$ yr$^{-1}$ remains.
Those solid circles at SSFR$_{\rm 321-1000Myr} > 10^{-9,5}$ yr$^{-1}$ correspond 
to selected PSBs. 
For PSBs, the median H$\delta_{\rm A}$ is $\sim 6$\AA\, and
14 out of 19 those galaxies have H$\delta_{\rm A} > 5$ \AA,
which is often used as a criterion for PSB or ``H$\delta$ strong'' galaxies. 
Therefore, we expect that many of our PSBs satisfy
the spectroscopic selection used in the previous studies, namely, 
relatively high H$\delta_{\rm A}$ and low EW([O{\footnotesize II}]),  
while our selection picks up those galaxies with quenching 
several hundreds Myr before observation and probably misses those with
more recent quenching within several tens to a hundred Myr.

In Figure \ref{fig:ssfrerr}, we show half widths of the 68\% confidence
intervals of SSFR$_{\rm 0-40Myr}$, SSFR$_{\rm 40-321Myr}$, and SSFR$_{\rm 321-1000Myr}$
as a function of SSFR itself.
In the middle panel, we separately plot the uncertainty of
SSFR$_{\rm 40-321Myr}$ for those galaxies with
SSFR$_{\rm 0-40Myr} < 10^{-10.5}$ yr$^{-1}$ with red circles as well as
all galaxies (magenta dots), because the uncertainty tends to be larger 
when younger population dominates in the SED and vice versa.
The uncertainty of SSFR$_{\rm 321-1000Myr}$ for 
those with SSFR$_{\rm 0-40Myr} < 10^{-10.5}$ yr$^{-1}$ and
SSFR$_{\rm 40-321Myr} < 10^{-10.5}$ yr$^{-1}$ are similarly shown as red circles
in the right panel.
In the left panel, the median uncertainty of SSFR$_{\rm 0-40Myr}$
is less than factor of 2 at
SSFR$_{\rm 0-40Myr} \gtrsim 10^{-10.5}$ yr$^{-1}$,
while the fractional error is larger 
at lower SSFRs.
The median uncertainty of SSFR$_{\rm 0-40Myr}$ is less than factor of 2 at
SSFR$_{\rm 0-40Myr} \gtrsim 10^{-10.5}$ yr$^{-1}$.
While the uncertainty of SSFR$_{\rm 40-321Myr}$ tends to be slightly larger
than that of SSFR$_{\rm 0-40Myr}$ in the middle panel,
galaxies with SSFR$_{\rm 0-40Myr} < 10^{-10.5}$ 
yr$^{-1}$ show the similar uncertainty at
SSFR$_{\rm 40-321Myr} \gtrsim 10^{-10}$ yr$^{-1}$, where the median value
is less than factor of 2.
The uncertainty of SSFR$_{\rm 321-1000Myr}$ is larger than those of
SSFR$_{\rm 0-40Myr}$
and SSFR$_{\rm 40-321Myr}$ (magenta dots in the right panel),
which reflects the fact that SED of older 
stellar population is more easily overwhelmed by those of younger
populations.
Nevertheless, we can constrain SSFR$_{\rm 321-1000Myr}$ within a factor
of 2 at SSFR$_{\rm 321-1000Myr} \gtrsim 10^{-10}$ yr$^{-1}$ for galaxies
with low SSFR$_{\rm 0-40Myr}$ and SSFR$_{\rm 40-321Myr}$ (red circles in the panel). 
\begin{figure}
  \includegraphics[width=\columnwidth]{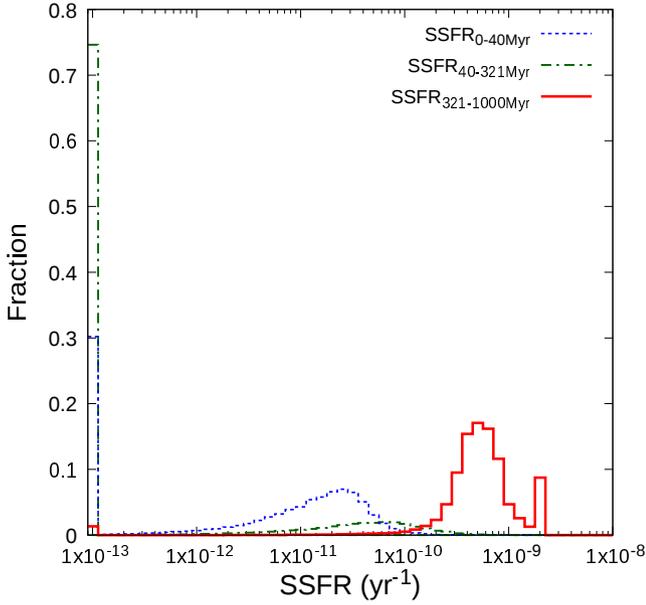}
  \caption{
    Average probability distributions of SSFR$_{\rm 0-40Myr}$ (dashed),
    SSFR$_{\rm 40-321Myr}$ (dashed-dotted), and
    SSFR$_{\rm 321-1000Myr}$ (solid) estimated in the SED fitting
    for 94 PSBs in our sample.
    The distributions for each PSB are calculated from the Monte Carlo
    simulations (see text for details),
    and those averaged over the 94 PSBs are shown.
    The probabilities of SSFR $< 10^{-13}$ yr$^{-1}$ are counted in a bin 
    at $10^{-13}$ yr$^{-1}$.
  \label{fig:pdf}}
\end{figure}

In Figure \ref{fig:pdf}, we show combined probability distributions of
SSFR$_{\rm 0-40Myr}$, SSFR$_{\rm 40-321Myr}$, and SSFR$_{\rm 321-1000Myr}$ for
94 PSBs to check whether these galaxies show significantly declining 
SFHs in the recent past even if the uncertainty is considered.
We averaged these probability distributions of SSFRs 
calculated from the Monte Carlo simulations described 
in the previous subsection 
over those PSBs with equal weight.  
One can see that the probability distribution of SSFR$_{\rm 321-1000Myr}$
is clearly higher than those of SSFR$_{\rm 0-40Myr}$ and SSFR$_{\rm 40-321Myr}$.
The probability of SSFR$_{\rm 321-1000Myr} > $2--3$ \times10^{-10}$ yr$^{-1}$
is high, 
while that of SSFR$_{\rm 40-321Myr} > 10^{-10}$ yr$^{-1}$ is very low 
(SSFR$_{\rm 40-321Myr} = 0$ in most cases). 
The distribution of SSFR$_{\rm 0-40Myr}$
has a peak around $2\times10^{-11}$ yr$^{-1}$, and 
there is only a very small probability 
of SSFR$_{\rm 0-40Myr} > 10^{-10}$ yr$^{-1}$.
Therefore, we expect that most of these PSBs had
experienced a high star formation activity
in 321--1000Myr before observation and then 
rapidly decreased their SFRs.

\subsubsection{Star-forming \& Quiescent Galaxies} \label{sec:comps}

\begin{figure}
  \includegraphics[width=\columnwidth]{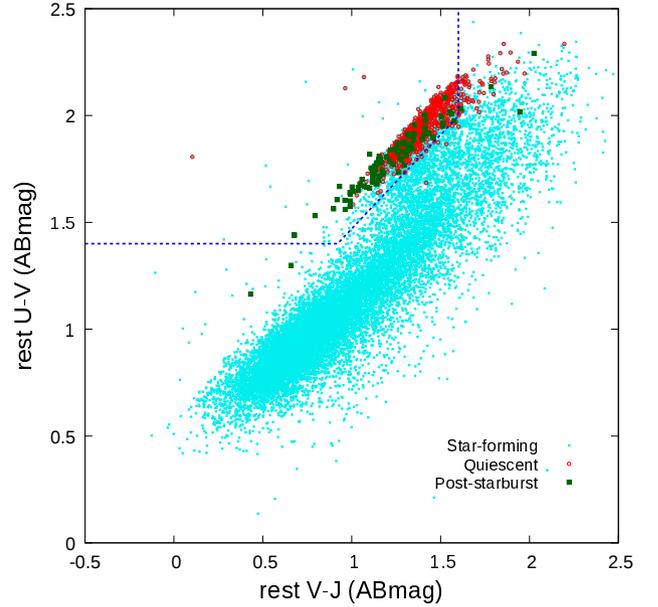}
  \caption{
    Rest-frame $U-V$ vs. $V-J$ two-colour diagram for SFGs (cyan dot),
    QGs (red open circle), and PSBs (green solid square) in our sample.
    The dashed line shows the colour criteria for QGs by \citet{will09}.
  \label{fig:uvj}}
\end{figure}

We also constructed comparison samples of normal SFGs  
and QGs in the same redshift range.
From those galaxies with $i<24$ and the reduced minimum $\chi^{2} < 5$ at
$0.7<z<0.9$, we selected those objects with
SSFR$_{\rm 0-40Myr} = 10^{-10}$--$10^{-9}$ yr$^{-1}$
and SSFR$_{\rm 40-321Myr} = 10^{-10}$--$10^{-9}$ yr$^{-1}$ as SFGs.
Since the both distributions of SSFR$_{\rm 0-40Myr}$ 
and SSFR$_{\rm 40-321Myr}$ show a peak around $10^{-9.5}$ yr$^{-1}$,
these galaxies are on and around the main sequence at least in
the last $\sim 300$ Myr.
We did not use SSFR$_{\rm 321-1000Myr}$ in the criteria for SFGs,
because the uncertainty of SSFR$_{\rm 321-1000Myr}$ tends to be 
large for those galaxies with relatively high SSFR$_{\rm 0-40Myr}$ and/or
SSFR$_{\rm 40-321Myr}$ (Figure \ref{fig:ssfrerr}).

For QGs, we selected galaxies with low SSFRs within
 recent 1 Gyr, namely, SSFR$_{\rm 0-40Myr} < 10^{-10.5}$ yr$^{-1}$ \&
SSFR$_{\rm 40-321Myr} < 10^{-10.5}$ yr$^{-1}$ \& 
SSFR$_{\rm 321-1000Myr} < 10^{-10.5}$ yr$^{-1}$.
We did not use those in the older periods than 1 Gyr in the criteria,
because it is difficult to strongly constrain detailed SFHs older than
1 Gyr due to the degeneracy mentioned above (Figure \ref{fig:temp}).
On the other hand, those QGs with a high
SSFR$_{\rm 1-2Gyr}$ could be similar with PSBs in that 
they had experienced a starburst at slightly earlier epoch than 
 PSBs followed by rapid quenching. 
Thus we check results for those galaxies in Section \ref{sec:casfh}.

Finally, we selected 6581 SFGs and 670 QGs
with $i<24$ and the reduced $\chi^{2} < 5$ at $0.7<z<0.9$.
Examples of the SED fitting for these galaxies are shown in Figure
\ref{fig:sedexam}.
In Figure \ref{fig:uvj}, we plot these SFGs, QGs, and
PSBs in the rest-frame $U-V$ vs. $V-J$ two-colour plane
in order to compare our classification with those in previous studies.
The rest-frame colours were estimated from the best-fit model templates
in the SED fitting.
The dashed line in the figure shows the criteria for QGs 
by \citet{will09}.
One can see that most of QGs and PSBs satisfy
the criteria, while some galaxies show redder $U-V$ and $V-J$ colours
due to relatively large dust extinction.
PSBs tend to have bluer colours than QGs,
which suggests younger stellar ages and more recent quenching of star formation.
If we use the rotated system of coordinates introduced by \citet{bel19} on
the two-colour diagram,  
our PSBs show $S_{\rm Q} = $ 1.60--2.55, while QGs have $S_{\rm Q} = $ 2.05--2.65.
The range of $S_{\rm Q}$ for our PSBs are similar with that of the
spectroscopic sample at $z=$ 1.0--1.5 in \citet{bel19}.
Such distribution in the two-colour plane is consistent with those in
previous studies of PSBs 
(e.g., \citealp{wil14}; \citealp{wil20}; \citealp{deu20};
\citealp{wu20}).
On the other hand, most SFGs are outside of the selection
area in the plane, although a small fraction of those galaxies
enter the area near the boundary.
Thus our classification with the SED fitting is consistent
with those in the previous studies, while our PSB selection
could miss those with very recent quenching of star formation, for
example, within $\sim$ 100 Myr.

\subsection{Morphological analysis} \label{sec:morp}

\subsubsection{Preparation}
  
In this study, we used three non-parametric morphological indices, namely,
concentration $C$, asymmetry $A$, and concentration of asymmetric features 
$C_{A}$,
to investigate morphological properties of PSBs.
We measured these indices on the {\it HST}/ACS $I_{\rm F814W}$-band images, while
we defined pixels that belong to the object in the 
Subaru/Suprime-Cam $i^{'}$-band data in order to keep consistency with
the object detection and SED analysis carried out with the ground-based data,
and to include discrete features/substructures such as knots, tidal tails,
and so on in the analysis.

We cut a 12'' $\times$ 12'' region centred on the object coordinate of
the ACS and Suprime-Cam data for each galaxy.
At first, we ran SExtractor version 2.5.0 \citep{ber96} on the $i^{'}$-band
images by using the RMS maps \citep{cap07} to scale detection threshold.
The detection threshold of 0.6 times RMS values from the RMS map  
over 25 connected pixels was used. 
We aligned segmentation maps output by SExtractor 
to the ACS $I_{\rm F814W}$-band images
with a smaller pixel scale, and used these maps to identify pixels
that belong to the object in the $I_{\rm F814W}$-band images.

While we basically used pixels identified by the segmentation map from
the $i^{'}$-band data, we added and excluded some pixels that belong to
objects/substructures detected in the $I_{\rm F814W}$-band images  
across the boundary defined by the segmentation map.
For this adjustment, we also ran SExtractor on the $I_{\rm F814W}$-band images
with a detection threshold of 1.2 times local background root mean
square over 12
connected pixels. If more than half pixels of a source detected in
the $I_{\rm F814W}$-band data are included in the object region defined 
by the $i^{'}$-band segmentation map, we included all pixels of this source
in the analysis and added some pixels outside of the object region if exist.
On the other hand, if less than half pixels of the source is included,
we masked and excluded all pixels of this source from the analysis
as another object. 
We used the adjusted object region to measure the morphological indices
for each galaxy.

We estimated pixel-to-pixel background fluctuation in the 12'' $\times$ 12''
region of the $I_{\rm F814W}$-band data by masking the object region
and pixels that belong to the other objects with the segmentation map.
We then defined pixels higher than 1$\sigma$ value of the background
fluctuation in the object region as the object, 
and masked the other pixels less than 1$\sigma$ in the region.

\subsubsection{Concentration}

Following previous studies such as \citet{ken85} and \citet{ber00},
we measured the concentration index defined as $C = r_{80}/r_{20}$, 
where $r_{80}$ and $r_{20}$ are radii which contain 80\% and 20\% of
total flux of the object.
The total flux was estimated as a sum of the 
pixels higher than 1$\sigma$ value of the background fluctuation
in the object region.
Since the object region is defined by an extent of the object in
the $i^{'}$-band data, which is often wider than that
in the ACS $I_{\rm F814W}$-band
data, contribution from background noise higher
than the 1$\sigma$ value is non negligible.
In order to estimate this contribution, we made sky images by replacing
pixels in the object region with randomly selected ones that do not
belong to any objects in the 12'' $\times$ 12'' field.
All pixels outside the object region in the sky image were masked with
zero value.
We defined pixels higher than 2$\sigma$ in the object image as
object-dominated region, and masked pixels at the same coordinates with
this region in the sky image, because no biased contribution from noises
is expected for those pixels with high object fluxes.
We summed up pixels higher than 1$\sigma$ value in the sky image
except for the object-dominated region, and subtracted this from
the total flux of the object.

We then measured a growth curve with circular apertures centred at 
a flux-weighted mean position of the object pixels  
 to estimate $r_{80}$ and $r_{20}$.
In measurements of the growth curve, the similar subtraction of  
the background contribution was performed with the same sky image.
We made 20 sky images for each object and repeated the measurements of $C$,
and adopted their mean and standard deviation as $C$ and its uncertainty,
respectively.

\begin{figure*}
  \includegraphics[width=2.1\columnwidth]{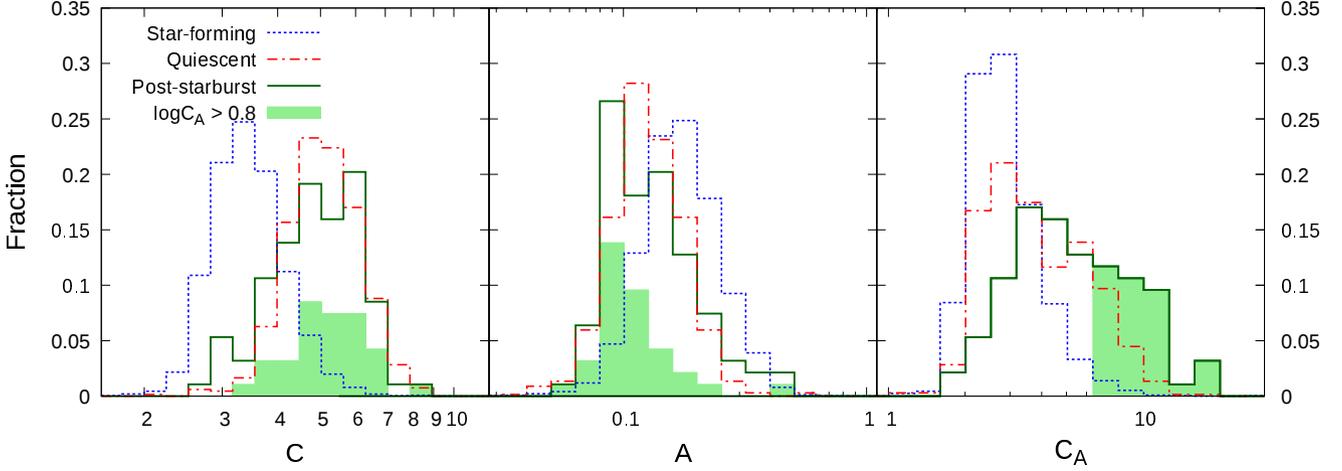}
  \caption{
    Distributions of the morphological indices, $C$, $A$, and $C_{A}$
    for SFGs (blue dashed), QGs (red dotted-dashed), and PSBs (green solid)
    in our sample. The light-green solid histogram represents the
    contribution from those PSBs with $\log{C_{A}} > 0.8$.
  \label{fig:histc}}
\end{figure*}

\subsubsection{Asymmetry}

We used the Asymmetry index defined by previous studies
(\citealp{sch95}; \citealp{abr96}; \citealp{con00}) to
measure rotation asymmetry of our sample galaxies.
The asymmetry index $A$ was calculated by rotating the object image
by 180 degree and subtracting it from the original image as
\begin{equation}
A = \frac{0.5 \Sigma |I_{\rm O} - I_{180}|}{\Sigma I_{\rm O}},  
\label{eq:asy}
\end{equation}
where $I_{\rm O}$ and $I_{180}$ are pixel values in the original image and
that rotated by 180 degree.

We adopted the same definition of the object pixels as in the calculation
of $C$, namely, those higher than 1$\sigma$ value in the object region.
The denominator of Equation (\ref{eq:asy}) is the same as the total flux of
the object described above, and the same correction for the background
contribution was applied.
In the calculation of the numerator, we chose a centre of the rotation 
so that a value of $A$ for the object is minimum, following \citet{con00}.
We searched such a centre with a step of 0.5 pixel, namely, 
centre of each pixel or boundary between two pixels in X and Y axes 
over the object region.
This step size is about six times smaller than the PSF FWHM of
$\sim$ 0.1 arcsec, and we can determine the rotation centre with
sufficiently high accuracy to avoid significant artificial residuals
in a central region due to errors of the centre.
After subtracting the rotated image from the original image, 
we masked negative pixel values in the residual image with zero value 
(hereafter rotation-subtracted image). 
We then summed (positive) fluxes in the object region of
the rotation-subtracted image.

By using the same sky image for the object as used in the calculation of
 total flux, 
we also estimated the contribution from the background noise for the
rotation-subtracted image as follows.
At first, we remained only pixels higher than 1$\sigma$ value in the sky image
and replaced the other pixels with zero.
Second, we rotated the masked sky image by 180 degree and subtracted it from
that before rotation.
We then summed up only positive pixels in the residual sky image except for
those pixels in the object-dominated region, where pixels in the object
residual image are higher than 2$\sigma$ value and the contribution
from asymmetric features of the object dominates.
Finally, we subtracted the estimated background contribution from
the summed flux of the object residual image.
By using 20 random sky images as in the estimate of $C$,
we repeated the calculations described above and
adopted the mean and standard deviation as $A$ and its uncertainty,
respectively.

\subsubsection{Concentration of Asymmetric Features}

We newly devised a morphological index measuring central concentration
of asymmetric features of the object, $C_{A}$.
This is a combination of the asymmetry and concentration indices described
above. With $C_{A}$, we aim to detect asymmetric features such as
central disturbances or tidal tails, and distinguish them from those
by star-forming regions in steadily star-forming disks.
For example, normal star-forming disk galaxies, which usually show 
a central bulge with little asymmetric feature and an extended disk
with star-forming regions at random positions, are expected to
have relatively low concentration of the asymmetric components in 
their surface brightness distribution.
While tidal tails tend to occur in outer regions of galaxies and
lead to lower concentration, nuclear starbursts induced by gas inflow
to the centre may cause disturbed features in their central region and
enhance this index.

We used the same rotation-subtracted object 
image as in the calculation of $A$. 
We similarly estimated radii which contain 80\% and 20\% of total flux
of the rotation-subtracted image, namely, $r_{A,80}$ and $r_{A,20}$,
and calculated the index as $C_{A} = r_{A,80}/r_{A,20}$.
To do this, we measured a growth curve on the rotation-subtracted image
with circular apertures centred at 
the rotation centre, for which the minimum value of $A$ was obtained.
The total flux of the residual image was the same as in the calculation of $A$,
where the contribution from the background noise was estimated and subtracted.
In measurements of $r_{A,80}$ and $r_{A,20}$, we similarly corrected for  
the contribution from the background noise 
with the same rotation-subtracted sky image.
We also repeated the calculations with 20 random sky images and
adopted the mean and standard deviation as $C_{A}$ and its uncertainty,
respectively.

\section{Results} \label{sec:res}

\begin{figure*}
  \includegraphics[width=2.1\columnwidth]{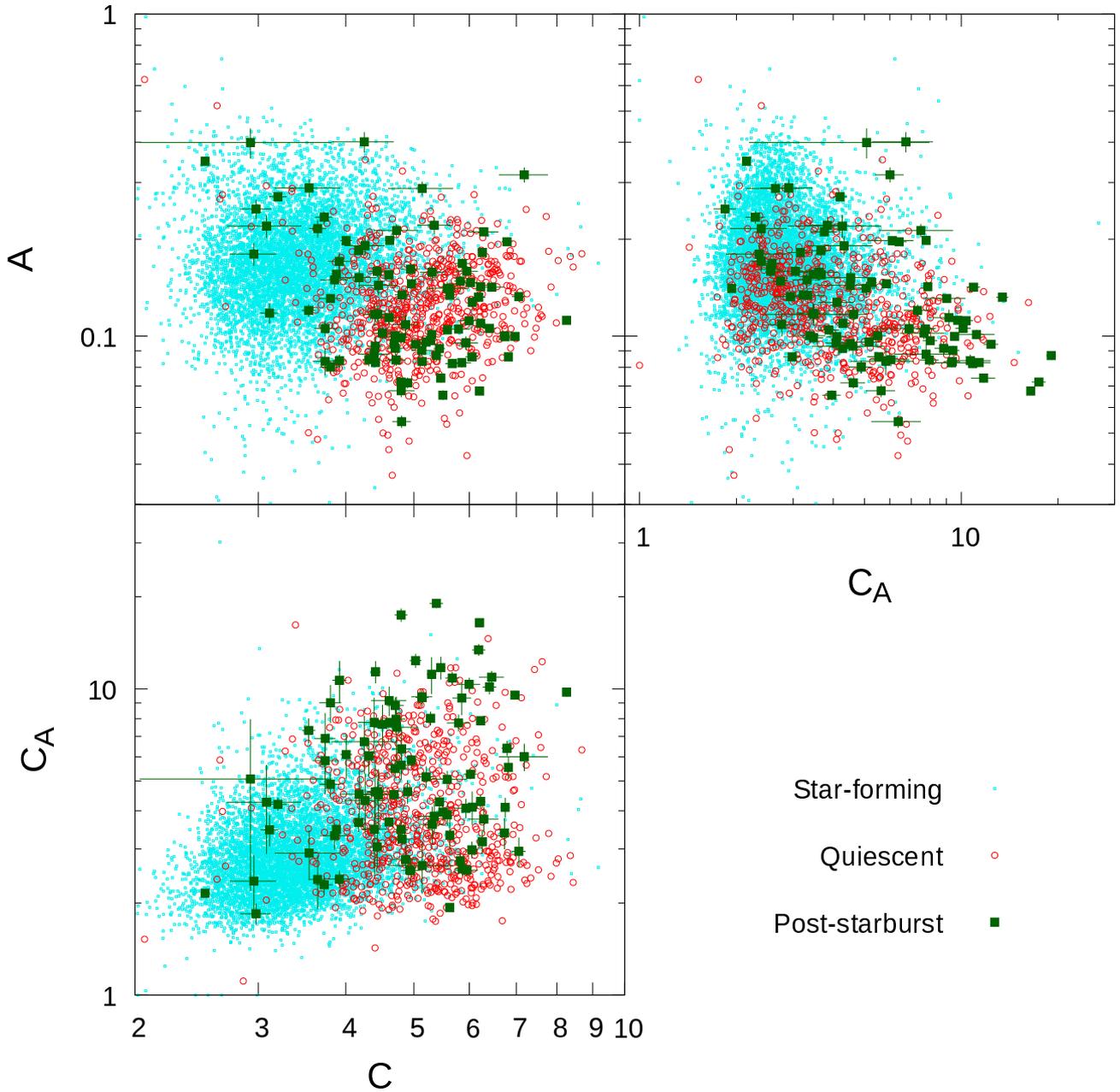}
  \caption{
    Scatter plots of the morphological indices, $C$, $A$, and $C_{A}$
    for SFGs (cyan dot), QGs (red open circle), and PSBs (green solid square)
    in our sample.
    The error bars for PSBs show the uncertainty of the indices calculated in
    Section \ref{sec:morp}, while those for SFGs and QGs are omitted for
    clarity.
 \label{fig:cacac}}
\end{figure*}

Figure \ref{fig:histc} shows distributions of $C$, $A$, and $C_{A}$
for PSBs with $i<24$ at $0.7<z<0.9$
in the COSMOS field.
For comparison, we also show SFGs and QGs described in
Section \ref{sec:comps}.
Most PSBs have $C\sim$ 3--7 and the $C$ distribution of PSBs is 
similar with that of QGs, while the fraction of PSBs with $C \lesssim$ 4
is slightly higher than QGs. PSBs show clearly higher $C$ values 
 than SFGs, most of which have $C\sim$ 2.5--4.5.
The $A$ distribution of PSBs shows a peak around $A\sim 0.09$ and most of
PSBs have $A\sim$ 0.06--0.3.  QGs show a similar range of $A$ and  
 a peak at a slightly higher value of 
$A \sim$ 0.12. SFGs have a higher distribution of $A$ with a peak around
$A \sim$ 0.15, and the fraction of SFGs with $A<0.1$ is small.
The $C$ and $A$ indices of PSBs are similar with QGs rather than SFGs,
while PSBs show slightly lower values for the both indices than QGs.

\begin{table*}
	\centering
	\caption{Median values of the morphological indices for the samples with the different SFHs \label{tab:cacac}}
	\label{tab:example_table}
	\begin{tabular}{ccccc} 
		\hline
		sample & $C$ & $A$ & $C_{A}$ & fraction of $\log{C_{A}}>0.8$\\
		\hline
		PSBs & $4.884\pm0.044$ & $0.1173\pm0.0032$ & $4.755\pm0.173$ & $0.362\pm0.019$\\
		QGs & $5.047\pm0.016$ & $0.1228\pm0.0007$ & $3.497\pm0.042$ & $0.158\pm0.007$\\
		SFGs & $3.396\pm0.003$ & $0.1681\pm0.0004$ & $2.723\pm0.007$ & $0.021\pm0.001$\\
		\hline
	\end{tabular}
\end{table*}

\begin{figure*}
  \includegraphics[width=2.1\columnwidth]{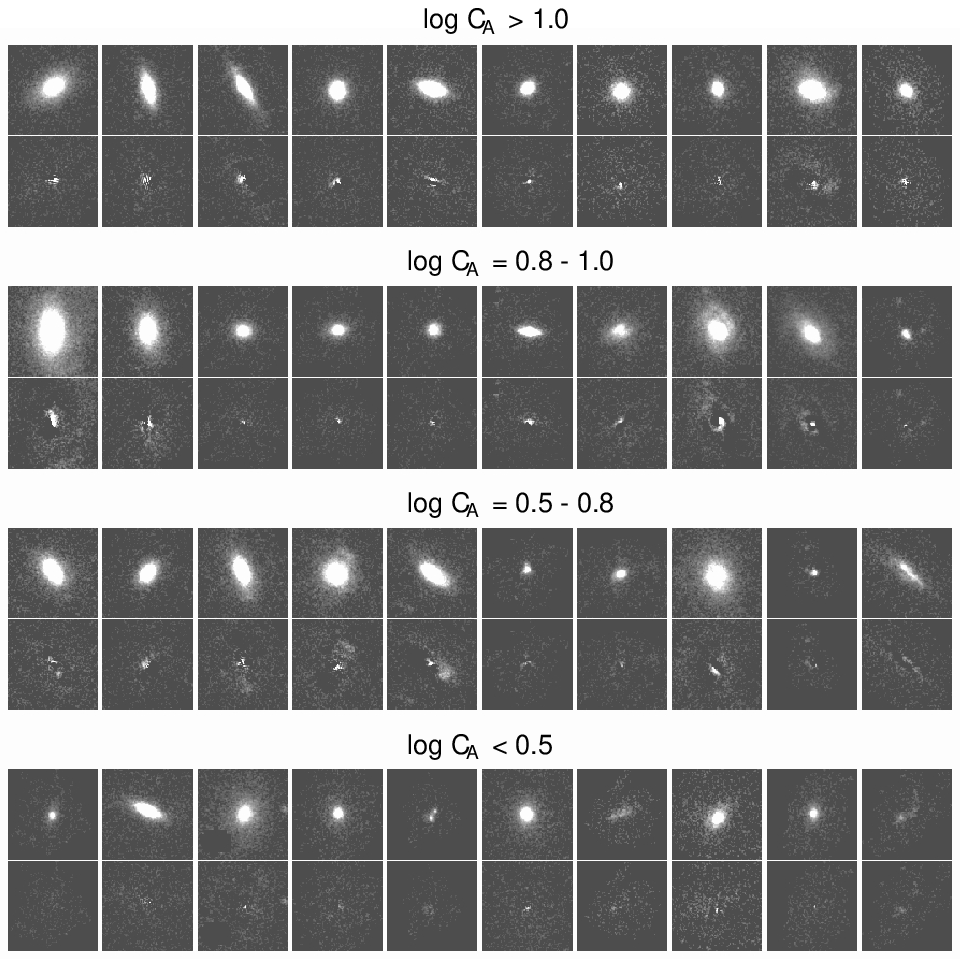}
  \caption{Examples of {\it HST}/ACS $I_{\rm F814W}$-band images for PSBs.
    Each panel is 3'' $\times$ 3'' in size. Two images are
    shown for each PSB: the upper panel shows the object image, and the lower
    panel shows the rotation-subtracted image. Four sets of the two rows of the
    images represent those with different $C_{A}$ values, and the $C_{A}$ values
    decease from top to bottom sets. In each set, galaxies are randomly
    selected from those with the $C_{A}$ values, and are sorted in
    order of increasing $A$. 
  \label{fig:mon}}
\end{figure*}

In contrast with these indices, PSBs clearly tend to show higher $C_{A}$
values than QGs. 
While  QGs show a peak around $C_{A} \sim 2.5$ and
a small fraction of those with $C_{A} \gtrsim 6$,
PSBs show a peak around $C_{A} \sim 4$ and more than a third of PSBs
have $C_{A} \gtrsim 6$.
SFGs have lower $C_{A}$ than QGs and most of them show $C_{A} < 6$.
We summarise median values of the morphological indices and the fraction
of those with high $C_{A}$ for PSBs, QGs, and SFGs in Table \ref{tab:cacac}.
The errors in Table \ref{tab:cacac} are estimated with Monte Carlo
simulations. In the simulations, we added random shifts
based on the estimated errors (Section \ref{sec:morp}) 
to the morphological indices of each galaxy in our sample,    
and calculated the median values of the
indices and the fraction of those with high $C_{A}$.
We repeated 1000 such simulations and adopted the standard deviations
of these values as the errors.
The median value of $C_{A} \sim 4.8$ for PSBs is significantly higher than
those of QGs and SFGs (3.5 and 2.7).
The fraction of PSBs with $\log{C_{A}} > 0.8$ is 36\%, while
those for QGs and SFGs are 16\% and 2\%, respectively.
A significant fraction of PSBs show the high concentration of the asymmetric
features.
We also performed a Kolmogorov-Smirnov test to statistically examine
the differences between PSBs and QGs. 
The probability that the $C_{A}$ distributions of PSBs and QGs are drawn from
the same probability distribution is only 0.003\%, while the probabilities
for $C$ and $A$ are 14.6\% and 15.8\%, respectively.

In Figure \ref{fig:cacac}, we show relationships among the morphological
indices for PSBs, QGs, and SFGs.
In the $C$ vs. $A$ panel, most of PSBs show relatively low $A$ and high $C$,
and their distribution is similar with that of QGs. There are also a 
small fraction of PSBs at relatively high $A$ and low $C$,
where most SFGs are located. 
In the $C_{A}$ vs. $A$ panel, $A$ tends to decrease with increasing $C_{A}$
for all the samples, while there is some scatter.
Most of PSBs with a high value of $C_{A} \gtrsim 6$ 
show $A \sim $ 0.07--0.15, while 
there are a few PSBs with $C_{A} \gtrsim 6$ and $A > 0.15$.
The $A$ distribution of those with $\log{C_{A}} >0.8$ is skewed to
lower values than that of all PSBs (Figure \ref{fig:histc}).
Those QGs with relatively high $C_{A}$ values show similarly low values of $A$.
In the $C_{A}$ vs. $C$ panel, the $C_{A}$ values of
galaxies with $C < 0.35$ are low, and the scatter of $C_{A}$ tends to
increase with increasing $C$. 
PSBs with a high value of  $C_{A} \gtrsim 6$ have $C \sim $ 3--8,
and their distribution of $C$ is not significantly different from
that of all PSBs.
Even if we limit to those with a very high value of $C_{A}>10$,
they have $C \sim $ 3.5--7, which is similar with those of QGs and
all PSBs.
Thus PSBs with high $C_{A}$ values tend to show similar $C$ and
slightly lower $A$ values than all PSBs and QGs.

\begin{figure*}
  \includegraphics[width=2.1\columnwidth]{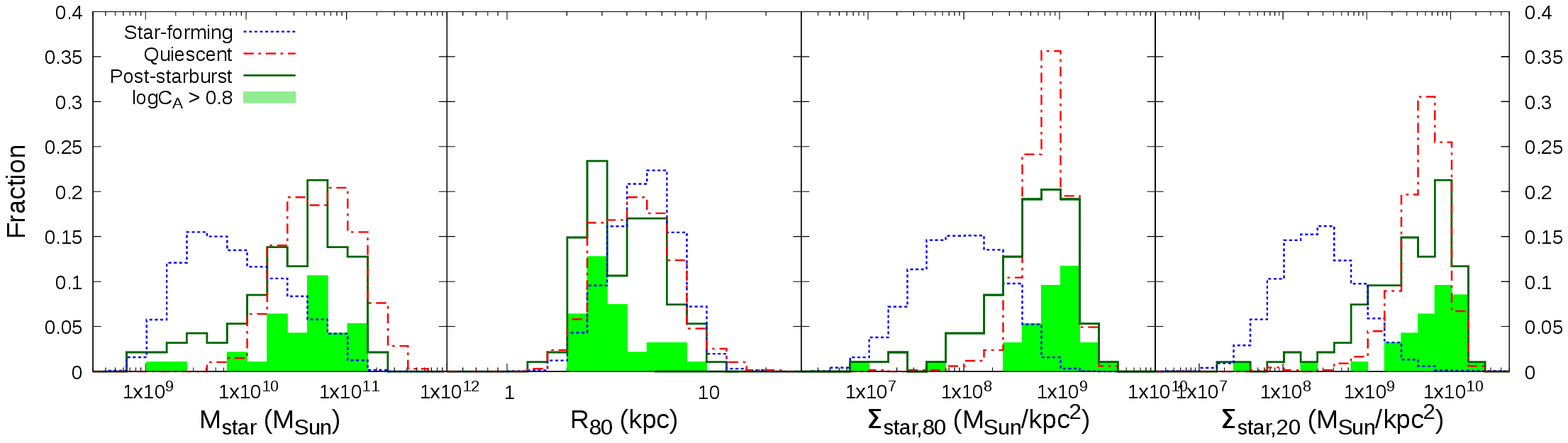}
  \caption{
    Distributions of stellar mass, $r_{80}$, 
    mean surface stellar mass density within $r_{80}$ ($\Sigma_{\rm star,80}$),
    and that within $r_{20}$ ($\Sigma_{\rm star,20}$) for SFGs, QGs, and PSBs
    in our sample. The symbols are the same as Figure \ref{fig:histc}.
    Note that the mean surface stellar mass densities are calculated under
    the assumption of a constant stellar M/L ratio over the entire galaxy. 
    \label{fig:msrsig}}
\end{figure*}

In Figure \ref{fig:mon}, we present examples of the ACS $I_{\rm F814W}$-band
object images and the rotation-subtracted residual ones for PSBs with
different $C_{A}$ values.
Many PSBs show early-type morphologies with a significant
bulge, while there are a few PSBs with low surface brightness and/or irregular
morphologies. 
One can see that PSBs with high $C_{A}$ values show  significant residuals
near their centre in the rotation-subtracted images.
While signal noises of the object could cause some residuals especially
in the central region where their surface brightness tends to be high,
the residuals show some extended structures 
near the centre rather than pixel-to-pixel fluctuations.
These asymmetric features seem to reflect physical properties in the central
region.

Those PSBs with high $C_{A}$ values 
tend to show relatively high surface brightness in the original images
and do not show large and bright asymmetric features such as tidal
tails in their outskirts.
On the other hand, those with low $C_{A}$ values have more extended 
asymmetric features in the rotation-subtracted images. 
Some of them show lower surface brightness and/or 
irregular morphologies in the object images.

\begin{figure*}
  \includegraphics[width=2.1\columnwidth]{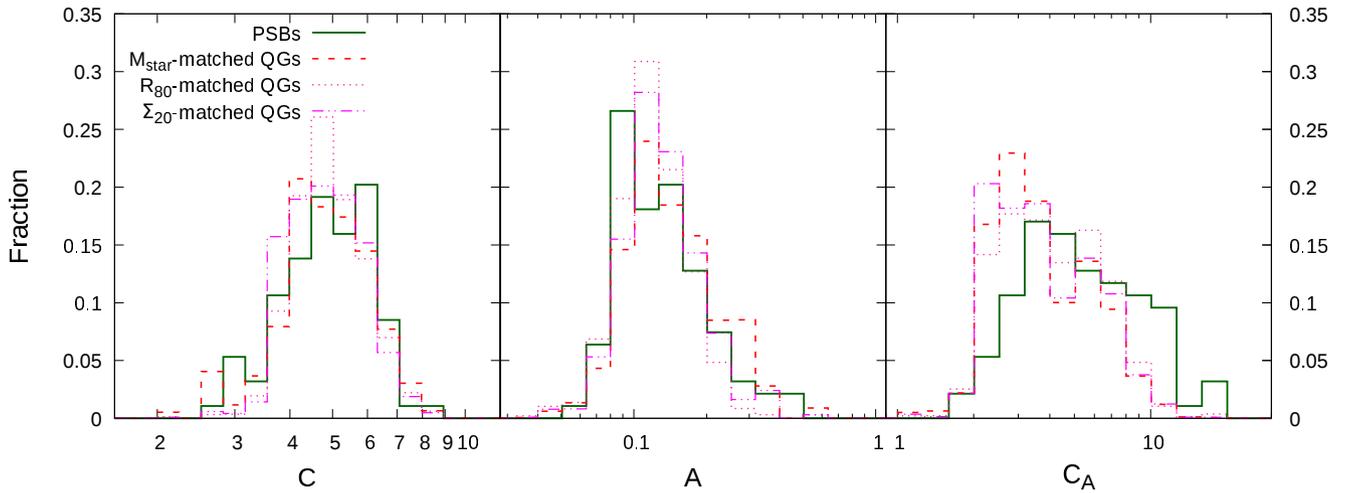}
  \caption{
    The same as Figure \ref{fig:histc} but for PSBs and control samples 
    of QGs. While the solid line shows PSBs, the dashed, dotted, and
    dashed-dotted
    lines represent $M_{\rm star}$, $r_{80}$, and $\Sigma_{20}$-matched samples
    of QGs, respectively.
    \label{fig:simqg}}
\end{figure*}

In Figure \ref{fig:msrsig}, we show $M_{\rm star, 0}$, $r_{80}$, and mean surface
stellar mass densities within $r_{80}$ and $r_{20}$ 
for all PSBs and those with $\log{C_{A}} > 0.8$.
Note that we assumed the surface stellar mass density has the same radial
profile with the $I_{\rm F814W}$-band surface brightness, and calculated 
 the mean surface stellar mass densities as
$\Sigma_{80(20)} = 0.8(0.2)\times M_{\rm star, 0} / (\pi r_{80(20)}^{2})$.
If the galaxy has centrally concentrated young stellar population,
which is often seen in PSBs as discussed in the next section,
we could overestimate 
the surface stellar mass density in the inner region, because the contribution
from the young population decreases    
 stellar mass to luminosity ratio in the central region.
While both all PSBs and those with high $C_{A}$ show similar or
slightly lower stellar 
mass distribution with a significant tail over $10^{9}$--$10^{10} M_{\odot}$
than QGs, PSBs tend to have smaller sizes ($r_{80}$) than QGs.
The fraction of relatively small galaxies with
$\log{(r_{80}/{\rm kpc})} <0.5$ in all PSBs (those with $\log{C_{A}} > 0.8$) 
is 42\% (53\%), which is larger than that of QGs (25\%). 
As a result, PSBs with $\log{C_{A}} > 0.8$ show higher $\Sigma_{\rm star, 80}$ 
values than QGs, while those of the other PSBs are similar or slightly
lower than QGs.
The higher surface stellar mass density of those PSBs with $\log{C_{A}} > 0.8$ than
QGs is more clearly seen in the inner region, i.e., $\Sigma_{\rm star, 20}$,
although we could overestimate these values as mentioned above.

In order to examine whether morphological properties of PSBs are
different from QGs or not when comparing galaxies with similar
physical properties such as stellar mass, size, and surface mass density,
we made control samples of QGs.
By using the distributions of stellar mass in Figure \ref{fig:msrsig} for
PSBs and QGs, we calculated ratios of fractions, namely, $f_{\rm PSB}/f_{\rm QG}$
in each stellar mass bin. We then used the ratio in each mass bin as
weight 
 to estimate distributions of the morphological indices for 
a sample of QGs that effectively has the same distribution of stellar mass
as PSBs.
We similarly estimated distributions of the indices for $r_{80}$-matched
and $\Sigma_{20}$-matched samples of QGs.
Figure \ref{fig:simqg} shows comparisons of the morphological indices
between PSBs and the control samples of QGs.
While the $M_{\rm star}$-matched sample shows slightly higher $A$ values,
the control samples basically have the similar distributions with
 the original sample of QGs.
 Even if we use the control samples of QGs, PSBs show significantly higher
 $C_{A}$ values than QGs, while their $C$ and $A$ values are similar with
 those of QGs.

\section{Discussion}

In this study, we fitted the photometric SEDs of galaxies with
$i<24$ from COSMOS2020 
catalogue with population synthesis models assuming 
non-parametric, piece-wise constant function of SFHs, and selected
94 PSBs with a high SSFR$_{\rm 321-1000Myr}$ ($>10^{-9.5}$ yr$^{-1}$) and low SSFR$_{\rm 40-321Myr}$ and SSFR$_{\rm 0-40Myr}$ ($< 10^{-10.5}$ yr$^{-1}$) at $0.7<z<0.9$.  
We measured the morphological indices, namely, $C$, $A$, and $C_{A}$ for
these PSBs on the ACS $I_{\rm F814W}$-band images and compared them with
SFGs whose SSFR$_{\rm 40-321Myr}$ and SSFR$_{\rm 0-40Myr}$ are on the main sequence
($10^{-10}$--$10^{-9}$ yr$^{-1}$) and QGs with low SSFR$_{\rm 321-1000Myr}$, SSFR$_{\rm 40-321Myr}$, and SSFR$_{\rm 0-40Myr}$ values ($< 10^{-10.5}$ yr$^{-1}$).
We found that PSBs show systematically higher $C_{A}$ than QGs and
SFGs, while their $C$ and $A$ are similar with those of QGs.
The fraction of PSBs with $\log{C_{A}} > 0.8$ is 36\%, which is significantly
higher than those of QGs and SFGs (16\% and 2\%).
We first examine relation between $C_{A}$ and SFHs to confirm how 
the $C_{A}$ values are related with their physical properties, 
and then discuss implications of the results for evolution of
these galaxies.

\subsection{Relation between $C_{A}$ and SFHs} \label{sec:casfh}

\begin{figure}
  \includegraphics[width=\columnwidth]{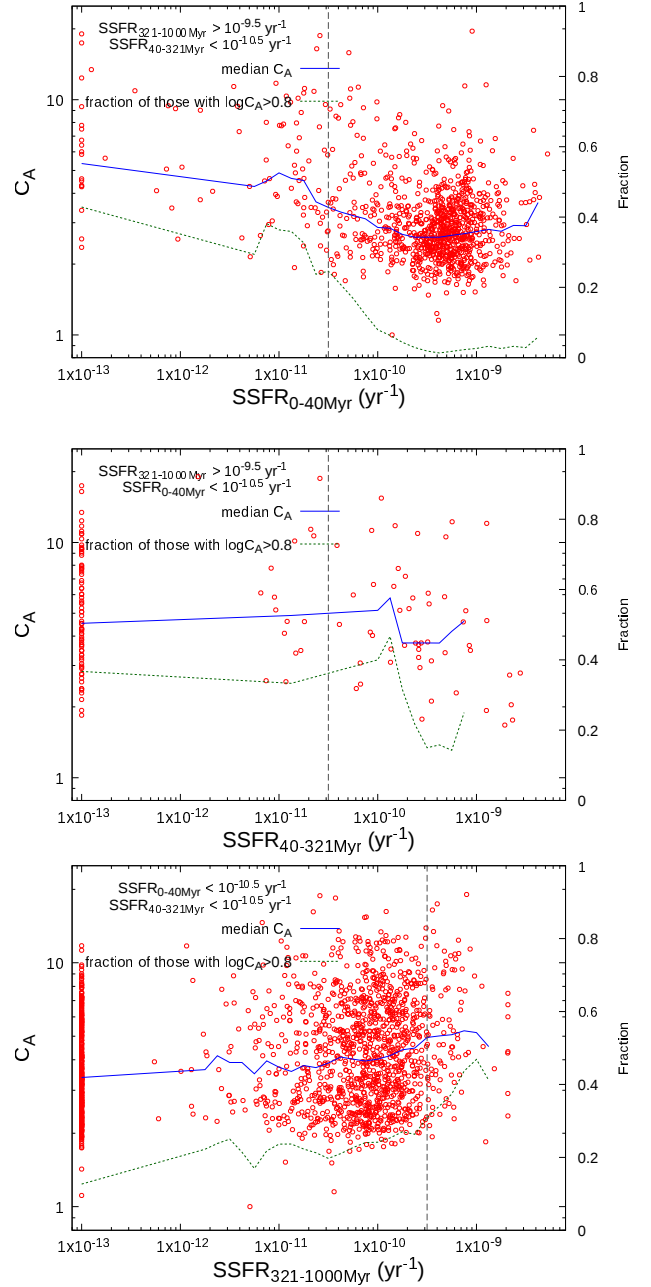}
  \caption{
    $C_{A}$ as a function of SSFR$_{\rm 0-40Myr}$ (top),
    SSFR$_{\rm 40-321Myr}$ (middle), and SSFR$_{\rm 321-1000Myr}$ (bottom)
    for PSBs and those galaxies that satisfy two out of the three
    SSFR criteria for the PSB selection (Equation (\ref{eq:psbsel})).
    For example, those with SSFR$_{\rm 321-1000Myr} > 10^{-9.5}$ yr$^{-1}$
    and SSFR$_{\rm 40-321Myr} < 10^{-10.5}$ yr$^{-1}$ are plotted in the top
    panel, and 
    those objects at SSFR$_{\rm 0-40Myr} < 10^{-10.5}$ yr$^{-1}$ are PSBs.
    Those with SSFR $ < 10^{-13}$ yr$^{-1}$ are plotted at $10^{-13}$ yr$^{-1}$.
    The solid line shows the median values of $C_{A}$ in SSFR bins with
    a width of $\pm$ 0.125 dex, while 
    the dashed line represents the fraction of those with $\log{C_{A}} > 0.8$.
  \label{fig:cass123}}
\end{figure}

We found that PSBs show higher $C_{A}$ than QGs and SFGs in the previous
section.
We here examine relationship between $C_{A}$ and SSFRs estimated
in the SED fitting.
The bottom panel of Figure \ref{fig:cass123} shows $C_{A}$ distribution
of those galaxies with SSFR$_{\rm 0-40Myr} < 10^{-10.5}$ yr$^{-1}$ and
SSFR$_{\rm 40-321Myr} < 10^{-10.5}$ yr$^{-1}$ as a function of SSFR$_{\rm 321-1000Myr}$.
Thus PSBs are located at SSFR$_{\rm 321-1000Myr} > 10^{-9.5}$ yr$^{-1}$
in this panel, while QGs are shown at SSFR$_{\rm 321-1000Myr} < 10^{-10.5}$ yr$^{-1}$.
The median value of $C_{A}$ and the fraction of those with $\log{C_{A}}>0.8$
clearly increase with increasing SSFR$_{\rm 321-1000Myr}$, especially at
SSFR$_{\rm 321-1000Myr} \gtrsim 10^{-10}$ yr$^{-1}$, while scatter at a given 
SSFR$_{\rm 321-1000Myr}$ is relatively large. 
This trend suggests that higher $C_{A}$ values of PSBs are related with 
their past star formation activities several hundreds Myr before observation.
We note that those with SSFR$_{\rm 321-1000Myr} \gtrsim 10^{-9}$ yr$^{-1}$ have 
slightly lower median values of $C_{A}$ and lower fraction of those with
high $C_{A}$. Those galaxies with very high SSFR$_{\rm 321-1000Myr}$ show 
relatively high asymmetry of $A \sim $ 0.15--0.40, and their asymmetric
features are widely distributed over the entire galaxies,
which could reduce their $C_{A}$ values.
Since SSFR$_{\rm 321-1000Myr} \sim  10^{-9}$ yr$^{-1}$ means that 
two thirds of the observed stellar mass had been formed in 321--1000 Myr
before observation (see the next subsection for details),
large morphological disturbances over the entire galaxies associated with  
such high star formation activities may remain at the observed epoch.  

\begin{figure*}
  \includegraphics[width=2.1\columnwidth]{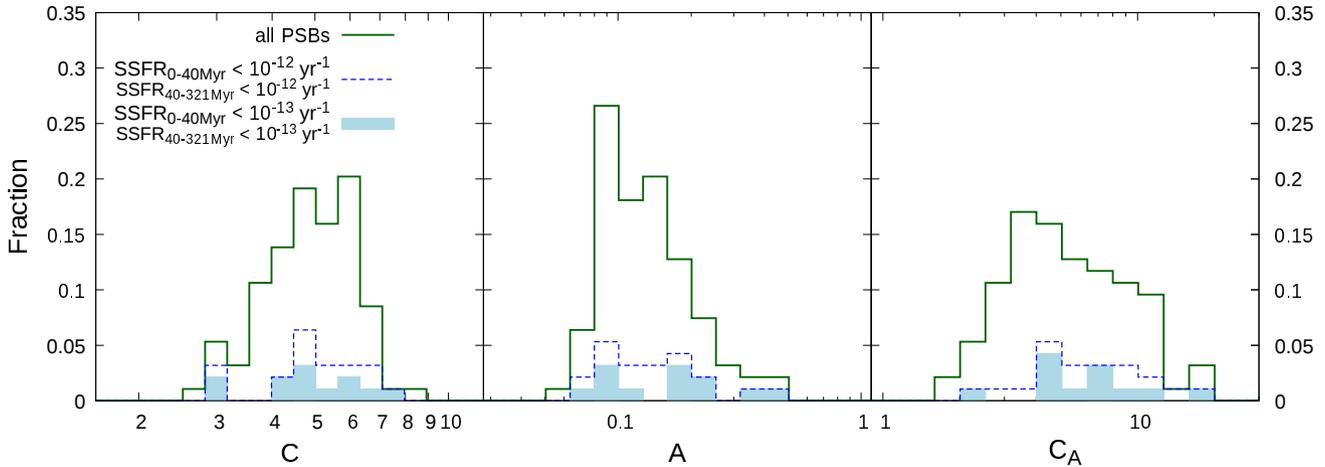}
  \caption{
    The same as Figure \ref{fig:histc} but for all PSBs and those with
    lower SSFR$_{\rm 0-40Myr}$ and SSFR$_{\rm 40-321Myr}$ values.
    While the solid line shows all PSBs, the dashed-line (solid histogram)
    represents contribution from those PSBs with
    SSFR$_{\rm 0-40Myr} < 10^{-12}$ yr$^{-1}$
    and SSFR$_{\rm 40-321Myr} < 10^{-12}$ yr$^{-1}$ 
    (SSFR$_{\rm 0-40Myr} < 10^{-13}$ yr$^{-1}$
    and SSFR$_{40-321Myr} < 10^{-13}$ yr$^{-1}$).
    \label{fig:psblow12}}
\end{figure*}

The top panel of Figure \ref{fig:cass123} shows the similar relationship
between $C_{A}$ and SSFR$_{\rm 0-40Myr}$ for those galaxies with
SSFR$_{\rm 321-1000Myr} > 10^{-9.5}$ yr$^{-1}$ and
SSFR$_{\rm 40-321Myr} < 10^{-10.5}$ yr$^{-1}$.
The median value of $C_{A}$ and the fraction of those with $\log{C_{A}}>0.8$
decrease with increasing SSFR$_{\rm 0-40Myr}$ at 
SSFR$_{\rm 0-40Myr} \gtrsim 10^{-11}$ yr$^{-1}$.
The relatively low values of $C_{A}$ for those galaxies
with SSFR$_{\rm 0-40Myr} = $ 10$^{-10}$--10$^{-9}$ yr$^{-1}$ on the main sequence
are probably caused by star-forming regions distributed over their disk,
while slightly higher $C_{A}$ values of those with SSFR$_{\rm 0-40Myr} > 10^{-9}$
yr$^{-1}$ may indicate some disturbances near the centre,
for example, by nuclear starburst.
The decrease of $C_{A}$ with increasing SSFR$_{\rm 0-40Myr}$ at
SSFR$_{\rm 0-40Myr} \lesssim 10^{-10}$ yr$^{-1}$ suggests that 
the high $C_{A}$ values seen in PSBs are not closely related with residual
on-going star formation.
The middle panel of Figure \ref{fig:cass123} shows the SSFR$_{\rm 40-321Myr}$
dependence of $C_{A}$, although most of those galaxies with 
SSFR$_{\rm 321-1000Myr} > 10^{-9.5}$ yr$^{-1}$ and
SSFR$_{\rm 0-40Myr} < 10^{-10.5}$ yr$^{-1}$ have SSFR$_{\rm 40-321Myr} = 0$ (plotted
at $10^{-13}$ yr$^{-1}$ in the figure).
One can see the similar trend that those with
SSFR$_{\rm 40-321Myr} = $ 10$^{-10}$--10$^{-9}$ yr$^{-1}$ 
show relatively low $C_{A}$ values.

\begin{figure}
  \includegraphics[width=\columnwidth]{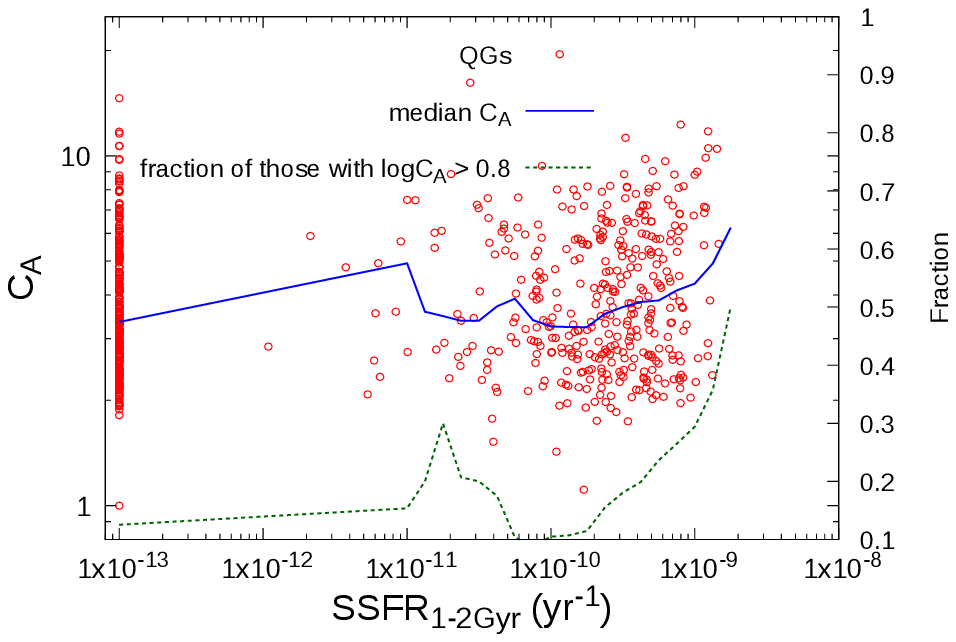}
  \caption{
    $C_{A}$ as a function of SSFR$_{\rm 1-2Gyr}$ for QGs, which satisfy
    all SSFR$_{\rm 0-40Myr}$, SSFR$_{\rm 40-321Myr}$, and SSFR$_{\rm 321-1000Myr}$
    are less than $10^{-10.5}$ yr$^{-1}$.
    The solid and dashed lines are the same as Figure \ref{fig:cass123}.
  \label{fig:cass4}}
\end{figure}

In Figure \ref{fig:psblow12}, 
we compare morphological indices for a sub-sample of PSBs with
lower values of SSFR$_{\rm 0-40Myr}$ and SSFR$_{\rm 40-321Myr}$,
namely, less than 10$^{-12}$ yr$^{-1}$ and 10$^{-13}$ yr$^{-1}$
with those of all PSBs.
Their distributions of the indices including $C_{A}$ are similar with
those of all PSBs.
This also suggests that residual star formation activities
within recent $\sim 300$ Myr in our PSBs do not seem to affect
their morphological properties significantly.

In Figure \ref{fig:cass4}, 
we also examined the $C_{A}$ distribution for QGs, which have low SSFRs
within the last 1 Gyr, as a function of SSFR$_{\rm 1-2Gyr}$, because 
16\% of these galaxies show $\log{C_{A}}>0.8$.
The median value of $C_{A}$ and the fraction of QGs with $\log{C_{A}}>0.8$ 
increase with increasing SSFR$_{\rm 1-2Gyr}$ at
SSFR$_{\rm 1-2Gyr} \gtrsim 10^{-9.5}$ yr$^{-1}$, in particular
at SSFR$_{\rm 1-2Gyr} > 10^{-9.0}$ yr$^{-1}$,
although the uncertainty of SSFR$_{\rm 1-2Gyr}$ is relatively large
as mentioned in Section \ref{sec:sample}.
Some of those QGs with a high SSFR$_{\rm 1-2Gyr}$ might have experienced   
a strong starburst at $\sim$ 1 Gyr before observation followed by
quenching of star formation, and similarly show significant asymmetric
features near the centre, which leads to high $C_{A}$ values.
Since about 40\% of QGs with $\log{C_{A}}>0.8$ have
SSFR$_{\rm 1-2Gyr} > 10^{-9.5}$ yr$^{-1}$, 
some fraction of the high $C_{A}$ values seen in QGs could also be related with
the past strong star formation activities.

\subsection {Implications for Quenching of Star Formation in PSBs}

We selected galaxies that experienced high star formation activities
followed by rapid quenching several hundreds Myr before observation,  
by using the SED fitting with the UV to MIR photometric data including
the optical intermediate-bands data.
We set the selection criteria to pick up those galaxies
whose SFR decreased by an 
order of magnitude between 321--1000 Myr and 40--321 Myr before
observation (Equation (\ref{eq:psbsel})).
Since we assumed a constant SFR in each period, 
the criterion of SSFR$_{\rm 321-1000Myr} > 10^{-9.5}$ yr$^{-1}$ means that
$\frac{SFR_{\rm 321-1000Myr}\times(1000-321)\times10^{6}}{M_{\rm star, 0}}>0.21$,
i.e., more than 20\% of the observed stellar mass had been
formed in the period of 321--1000 Myr before observation.
Thus PSBs in this study should have a strong starburst or 
rather high continuous star formation activities in the period.
Our selection could miss many of ``H$\delta$ strong'' galaxies
selected by previous studies, because those galaxies whose SFRs 
declined on a sufficiently short timescale (e.g., less than a few hundreds Myr)
could be H$\delta$ strong without such a strong starburst
(e.g., \citealp{leb06}; \citealp{paw18}).
In fact, if we limit our sample 
to those with $M_{\rm star, 0} > 10^{10} M_{\odot}$,
which roughly corresponds to the limiting stellar mass for our magnitude
limit of $i<24$ (e.g., \citealp{sat19}), 
the fraction and co-moving number density of PSBs become
$\sim$ 1\% (76/7266) and $3.5\times 10^{-5}$ Mpc$^{-3}$, respectively.  
These values are smaller than those of spectroscopically selected
(H$\delta$ strong) PSBs at similar redshifts in previous studies
(\citealp{wild09}; \citealp{yan09}; \citealp{ver10}; \citealp{row18}),
although the fraction
and number density of such PSBs depend on stellar mass, environment,
and selection criteria.
Our PSBs seem to have the similar burst strength and
time elapsed after the burst (several hundreds Myr) 
with those selected by photometric
SEDs (\citealp{wil16}; \citealp{wil20}). 
While the number density of PSBs with $M_{\rm star, 0} > 10^{10} M_{\odot}$
in our sample is still lower than that reported in \citet{wil16}, 
those of massive PSBs with $M_{\rm star} \gtrsim 10^{10.8} M_{\odot}$ are similar.
The fraction of PSBs
with $M_{\rm star, 0} \lesssim 10^{10} M_{\odot}$ is small in our sample
(the left panel of Figure \ref{fig:msrsig}), 
while \citet{wil16} found that the stellar mass function of those PSBs
at $z=$ 0.5--1.0 shows a steep low-mass slope. 
Thus we could miss such relatively low-mass PSBs probably due to the magnitude
limit of $i<24$.

In Figures \ref{fig:histc} and \ref{fig:cacac}, PSBs show relatively high
$C$ and low $A$ values, which are similar with those of QGs.
The high concentration in the surface brightness of PSBs is consistent
with results by previous studies in the local universe 
(e.g., \citealp{qui04}; \citealp{yam05}; \citealp{yan08}: \citealp{pra09}).
At intermediate redshifts, several studies also found that massive PSBs
with $M_{\rm star} \gtrsim 10^{10}M_{\odot}$ tend to have early-type morphology
with a high concentration, while there are many low-mass PSBs with
more disky morphology (\citealp{tra04}; \citealp{ver10}: \citealp{mal18}).
On the other hand, many previous studies reported that a significant fraction of
PSBs show asymmetric morphologies with tidal/disturbed features
(e.g., \citealp{zab96}; \citealp{bla04}; \citealp{tra04}; \citealp{yam05};
\citealp{yan08}; \citealp{pra09}; \citealp{won12}; \citealp{deu20}),
which seems to be
inconsistent with the low $A$ values of PSBs in our sample.
The discrepancy may be explained by differences in the time elapsed after
 starburst (hereafter, burst age).
\citet{paw16} found that the fraction of local PSBs
with disturbed features decreases
with increasing burst age, and their $A$ and shape 
asymmetry $A_{S}$, which is expected to be more sensitive to faint
tidal/disturbed 
features at outskirt,  become similar with 
those of normal galaxies at $\sim 0.3$ Gyr after the burst.
\citet{saz21} reported the similar anti-correlation between $A$ and burst
age for CO-detected PSBs in the local universe.
Theoretical studies with numerical simulations  
also suggest that such tidal/disturbed features in gas-rich major mergers
weaken with time and disappear within $\sim$ 0.1--0.5 Gyr after 
coalescence (\citealp{lot08}; \citealp{lot10}; \citealp{sny15}; \citealp{paw18};
\citealp{nev19}).
Since our selection method picks up those PSBs that experience a
starburst several hundreds Myr before observation
followed by quenching, their asymmetric features could already 
 weaken or disappear at the observed epoch.

In this study, we found that PSBs show higher $C_{A}$ values than QGs
and SFGs (Figures \ref{fig:histc} and \ref{fig:cacac}).
The high $C_{A}$ values are caused by the existence of significant
asymmetric features near the centre (Figure \ref{fig:mon}).
Such asymmetric features near the centre of PSBs in the local universe
have been reported by several studies.
\citet{yan08} measured $A$ of 21 PSBs with strong Balmer absorption lines
and weak/no [O{\footnotesize II}] emission varying aperture size, and found
that the $A$ values of most PSBs increase with decreasing aperture size,
which is different from those of normal spiral galaxies.
\citet{yam05} studied colour maps of 22 PSBs with strong H$\delta$
absorption and no significant H$\alpha$ emission at $z<0.2$, and found that 
some of them
show blue asymmetric/clumpy features near the centre in their colour maps.
\citet{pra09} also reported the similar irregular colour structures in the
central region for two out of 10 PSBs at $z<0.2$.
\citet{saz21} analysed {\it HST}/WFC3 data of 26 CO-detected
PSBs at $z<0.2$ and  
measured their morphological parameters such as
$A$, $A_{S}$, and residual flux fraction (RFF), which is the fraction of
 residual flux after subtracting best-fitted smooth S\'ersic profile.
They found that those with older burst ages tend to show
relatively low $A_{S}$ and high $A$ and RFF, which suggests that
internal disturbances could continue for longer time than tidal features
in outer regions.
For PSBs with high $C_{A}$ values in our sample, typical $r_{A,20}$ values
are $\sim$ 0.08--0.13 arcsec, which corresponds to $\sim$ 0.6--1.0 kpc
for galaxies at $z\sim 0.8$.
Therefore, the inner asymmetric features of those galaxies  
are located within $\sim 1$ kpc from the centre, and the spatial resolution
of $\lesssim $ 1 kpc seems to be required to reveal such high concentration
of the asymmetric features.

The existence of such asymmetric features near the centre
in $\sim$ 36\% of our PSBs suggests that disturbances in the
central region are closely related with rapid quenching of star formation.
One possible scenario is that such disturbances near the centre are
associated with nuclear starbursts, which occur in the galaxy mergers or
accretion events and lead to the quenching through rapid gas consumption
and/or gas loss/heating by supernova explosion, AGN outflow, tidal force,
and so on (e.g., \citealp{bek05}; \citealp{sny11}; \citealp{dav19}).
Several previous studies investigated radial gradients of Balmer
absorption lines and colours, and found that PSBs tend to show
the stronger Balmer absorption lines and bluer colours in the inner regions
(\citealp{yam05}; \citealp{yan08}; \citealp{pra13};
\citealp{che19}; \citealp{deu20}).
These results suggest that stronger starbursts occurred 
in the central region of these PSBs.
The numerical simulations of gas-rich galaxy mergers also predict that 
strong nuclear starbursts lead to the strong Balmer absorption lines and
blue colours in the central region of PSBs 
(\citealp{bek05}; \citealp{sny11}; \citealp{zhe20}).
In the nuclear starburst, stars are formed from kinematically disturbed
infalling gas, and spatial distribution and kinematics of newly formed
stars in the central region could also have disturbed/asymmetric features.
Remaining dust in the central region after the burst could cause
morphological disturbances in PSBs (e.g., \citealp{yan08}; \citealp{lot08};
\citealp{saz21}; \citealp{sme22}).
Since the fraction of PSBs with high $C_{A}$ in our sample increases with
increasing SSFR$_{\rm 321-1000Myr}$ (Figure \ref{fig:cass123}), 
their high $C_{A}$ values could be caused by such nuclear starburst 
several hundreds Myr before observation.
Those PSBs with $\log{C_{A}} > 0.8$ have the best-fit 
$A_{V} \lesssim$ 1.0 mag (median $A_{V} = 0.5$ mag) and
$[3.6] - [4.5] \sim $ 0, which are similar with or slightly lower and
bluer than those of the other PSBs. 
While most of them are not heavily obscured by dust over
the entire galaxies,  
some of those with high $C_{A}$ show dust-lane like features 
along the major axis of their surface brightness distribution in 
the rotation-subtracted images (Figure \ref{fig:mon}). 
The remaining dust in the central region could cause the asymmetric
features in their morphology.
Since several studies suggest that molecular gas and dust masses
in PSBs at low redshifts 
decrease by $\sim $ 1 dex in $\sim$ 500--600 Myr after starburst
(\citealp{row15}; \citealp{fre18}; \citealp{li19}), a significant
fraction of our PSBs, which are expected to experience a starburst 
several hundreds Myr before observation, could have the remaining gas
and dust in their central region.
The relatively high $C_{A}$ values of PSBs in our sample may indicate that 
disturbances in the stellar and/or dust distribution near the centre
tend to sustain for longer time after (nuclear) starburst 
than asymmetric features in outer regions such as tidal tails.

We found that PSBs, especially, those with high $C_{A}$ values
tend to have smaller sizes and higher surface stellar mass densities than
QGs (Figure \ref{fig:msrsig}).
The similar smaller sizes at a given stellar mass for PSBs at low and
intermediate redshifts have been  
reported by previous studies (\citealp{mal18}; \citealp{wu18};
\citealp{set22}; \citealp{che22}).
These results are consistent with the scenario where nuclear starburst 
causes asymmetric features 
near the centre in those PSBs with high $C_{A}$.
The nuclear starburst is expected to increase the stellar mass density
in the central region, which leads to decreases in half-light
and half-mass radii of these galaxies (e.g., \citealp{wu20}).
If this is the case, 
the half-light radii of those PSBs will   
gradually increase and become similar with those of QGs, because 
flux contribution from the young population in the central region  
is expected to decrease as time elapses.

Although the results in this study suggest the relationship between
nuclear starburst and quenching of star formation in those PSBs, 
we cannot specify why their star formation rapidly declined and 
has been suppressed for (at least) a few hundreds Myr.
Further observations of these galaxies will allow us
to reveal the quenching process. 
NIR and FIR imaging data with similarly high spatial resolution of
$\lesssim 1$ kpc scale taken with JWST and ALMA enable to investigate
details of the asymmetric features near the centre in those PSBs and
their origins. Physical state of molecular gas over the entire galaxies
is also important to understand the quenching mechanism(s).
Observing stellar absorption lines and nebular emission lines 
by deep optical (spatially resolved) spectroscopy allows us to 
study detailed stellar population and excitation state of ionised
gas. Although we excluded AGNs from the sample in this study
because of the difficulty of estimating non-parametric SFHs
with the AGN contribution in the SED fitting, it is interesting to
investigate relationship between the disturbances in the central region
and AGN activity.

\section{Summary}

In order to investigate morphological properties of PSBs,
we performed the SED fitting 
with the UV--MIR photometric data including the optical intermediate bands 
 for objects with $i<24$ from COSMOS2020 catalogue,
and selected 94 PSBs that experienced 
a high star formation activity several hundreds Myr before observation 
(SSFR$_{\rm 321-1000Myr} > 10^{-9.5}$ yr$^{-1}$)  followed by quenching 
(SSFR$_{\rm 40-321Myr} < 10^{-10.5}$ yr$^{-1}$ and SSFR$_{\rm 0-40Myr} < 10^{-10.5}$
yr$^{-1}$) at $0.7<z<0.9$.
We measured the morphological indices, namely, concentration
$C$, asymmetry $A$, and concentration of asymmetric features $C_{A}$,
on the {\it HST}/ACS $I_{\rm F814W}$-band images, and compared them with those
of QGs and SFGs. 
Our main results are summarised as follows.
\begin{itemize}

\item PSBs show relatively high concentration (the median  
  $C \sim 4.9$) and low asymmetry (the median $A \sim 0.12$),
 which are similar with those of QGs rather than SFGs. 
 Our selection method, which preferentially picks up those with 
 relatively old burst ages of several hundreds Myr, could lead to 
 the low asymmetry.

\item PSBs tend to show higher $C_{A}$ values (the median $C_{A} \sim 4.8$)
  than both QGs and SFGs (the median $C_{A} \sim 3.5$ and 2.7).
  The difference of $C_{A}$ between PSBs and QGs is significant
  even if we compare PSBs with $M_{\rm star}$, $r_{80}$, or $\Sigma_{20}$-matched
  samples of QGs.
  The fraction of galaxies with $\log{C_{A}} > 0.8$ in PSBs is $\sim $36\%,
  which is much higher than those of QGs and SFGs (16\% and 2\%).
  Those PSBs with high $C_{A}$ show remarkable asymmetric features
  near the centre,
  while they have relatively low overall asymmetry ($A \sim 0.1$).

\item The fraction of those PSBs with $\log{C_{A}} > 0.8$ increases
  with increasing SSFR$_{\rm 321-1000Myr}$ and decreasing SSFR$_{\rm 0-40Myr}$,
  which indicates that the asymmetric features near the centre are closely
  related with the high star formation activities several hundreds Myr
  before observation rather than residual on-going star formation.
  
\item Those PSBs with high $C_{A}$ tend to have higher surface
  stellar mass density, in particular, in the central region (e.g.,
  $\lesssim 1$ kpc) than both QGs and the other PSBs, while most of them 
  have the similar stellar masses of
  $1 \times 10^{10}$--$2 \times 10^{11} M_{\odot}$.

\end{itemize}  

These results suggest that a significant fraction of PSBs experienced 
 nuclear starburst in the recent past, and the quenching of 
star formation in these galaxies could be related with such active
star formation in the central region.
The high $C_{A}$ values of PSBs may indicate that the disturbances near
the centre tend to sustain for longer time than those at outskirt such as
tidal tails. If this is the case, $C_{A}$ could be used as morphological
signs of the past nuclear starburst in those galaxies with relatively 
old burst ages.

\section*{Acknowledgements}

We thank the anonymous referee for the valuable suggestions and comments.
This research is based in part on data collected at Subaru Telescope,
which is operated by the National Astronomical Observatory of Japan.
We are honoured and grateful for the opportunity of observing the 
Universe from Maunakea, which has the cultural, historical and natural 
significance in Hawaii.
Based on data products from observations made with ESO Telescopes at the La Silla Paranal Observatory under programme IDs 194A.2005 and 1100.A-0949(The LEGA-C Public Spectroscopy Survey). The LEGA-C project has received funding from the European Research Council (ERC) under the European Unions Horizon 2020 research and innovation programme (grant agreement No. 683184).
Data analysis were in part carried out on common use data analysis computer
system at the Astronomy Data Center, ADC, of the National Astronomical
Observatory of Japan.

\section*{Data Availability}

The COSMOS2020 catalogue is publicly available at
https://cosmos2020.calet.org/.
The COSMOS {\it HST}/ACS $I_{\rm F814W}$-band mosaic data version 2.0
are publicly available via NASA/IPAC Infrared Science Archive at
https://irsa.ipac.caltech.edu/data/COSMOS/images/acs\_mosaic\_2.0/.
The raw data for the ACS mosaic are available via
Mikulski Archive for Space Telescopes at
https://archive.stsci.edu/missions-and-data/hst.
The Subaru/Suprime-Cam $i{'}$-band mosaic reduced data
are also publicly available 
at https://irsa.ipac.caltech.edu/data/COSMOS/images/subaru/.
The raw data for the Suprime-Cam mosaic are accessible through
Subaru Telescope Archive System at 
https://stars.naoj.org.
The zCOSMOS spectroscopic redshift catalogue is publicly available via
ESO Science Archive Facility at
https://www.eso.org/qi/catalog/show/65.
The LEGA-C catalogue is also publicly available at 
https://www.eso.org/qi/catalogQuery/index/379.








\appendix

\section{Linear independence of templates}

\begin{figure}
  \includegraphics[width=\columnwidth]{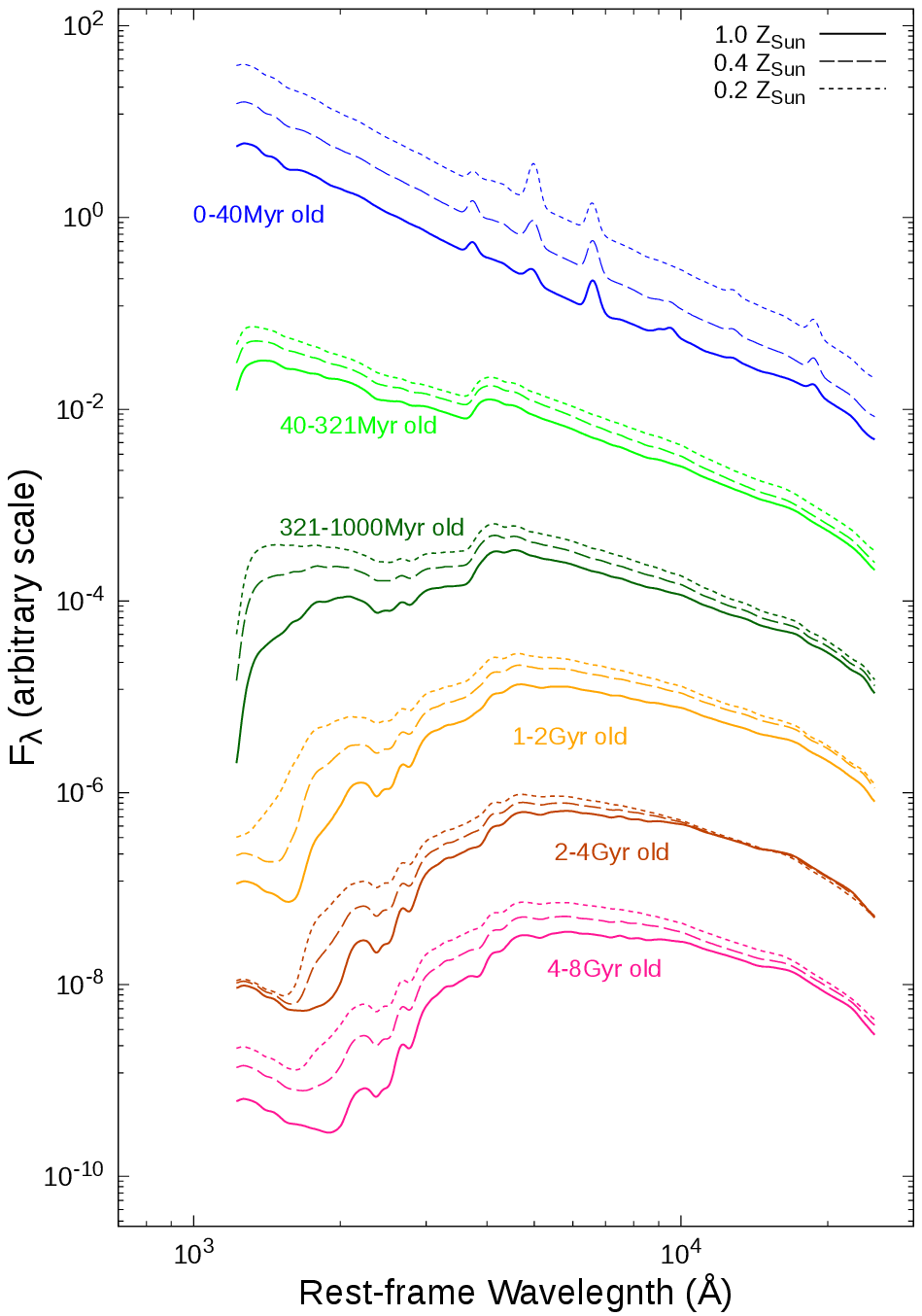}
  \caption{
    The same as Figure \ref{fig:temp}, but for those convolved with a Gaussian function with FWHM of $\lambda/20\AA$, which corresponds to the wavelength
    resolution of the intermediate bands.    
    \label{fig:convtemp}}
\end{figure}

\begin{figure}
  \includegraphics[width=0.95\columnwidth]{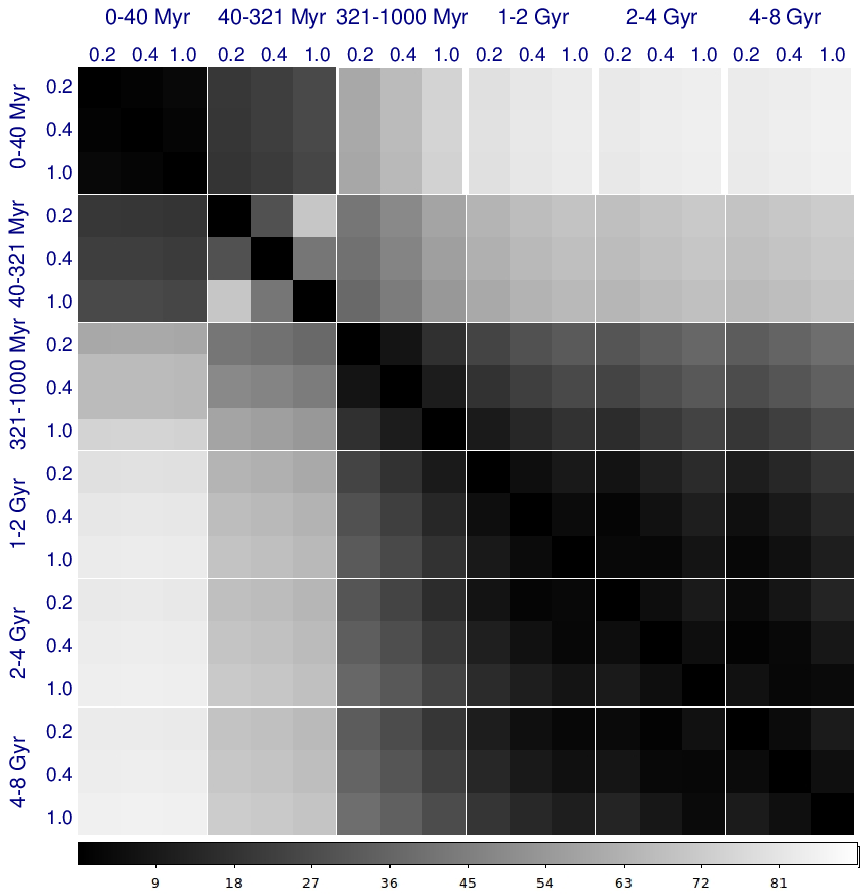}
  \caption{
    The angle $\theta_{ij}$ (Equation (\ref{eq:angle})) between spectral templates of stars formed in the different periods of look-back time. Results for three stellar metallicities,
    namely, 0.2, 0.4, and 1.0 $Z_{\odot}$ are shown for each period.       
    \label{fig:iptheta}}
\end{figure}

We here examine the linear independence of the model templates of stars formed
in the different periods of look-back time, which were used in the SED fitting.
Following \citet{mag15}, we considered the model templates as vectors, and
calculated inner (dot)  products between two of these vectors for the purpose.
At first, we convolved these templates with a spectral resolution of
the intermediate bands, namely, $\lambda/\Delta\lambda = 20$
(Figure \ref{fig:convtemp}), because 
we used the photometric SEDs in the SED fitting as described in
Section \ref{sec:sed}.
The 12 intermediate bands cover only $\sim$ 4000--8000 \AA,
which corresponds to the rest-frame 2000--5000 \AA\  at $z=$ 0.7--0.9,
and the wavelength resolution becomes lower at longer wavelengths.
While we here assume the resolution of $\lambda/\Delta\lambda = 20$ in all
wavelengths for simplicity, this does not affect results described below.
We used the template of 0--40 Myr where 50\% of ionising photons from stars 
really ionise gas and contribute to the nebular emission, 
but following results do not strongly depend on this fraction. 
We then calculate 'angle' between the templates $T_{i}$ and $T_{j}$ as 
\begin{equation}
\theta_{ij} = cos^{-1}\biggl(\frac{\sum\limits_{m}T_{i}(\lambda_{m})T_{j}(\lambda_{m})}{\sqrt{\sum\limits_{m}T_{i}(\lambda_{m})^{2}}\sqrt{\sum\limits_{m}T_{j}(\lambda_{m})^{2}}}\biggl), 
  \label{eq:angle}
\end{equation}
where $\lambda_{m}$ is wavelength of each resolution element.

Figure \ref{fig:iptheta} shows the angles between templates with different
periods and metallicities.
The angles between young ($<$ 321 Myr) templates and older ($>$ 1 Gyr) ones
are $\sim 90^\circ$, while those among the older templates are nearly zero.
The templates of 321--1000 Myr show intermediate angles with those of the other
periods, depending on metallicity.
In particular, the angle between a template of 321--1000 Myr with high
metallicity and 1--2 Gyr one with low metallicity is relatively low.
While this could cause some degeneracy, we expect that this does not
strongly affect our results because we assume the fixed metallicity
except for the youngest period.
Those of the same periods with different metallicities
show small angles except for 40--321 Myr ones, and effects of the different
metallicities seem to be not so large.
Thus we can basically differentiates contributions from the young, intermediate
(321--1000 Myr), old stellar populations, while the degeneracy could affect
our selection for PSBs if the metallicity of stars formed in starburst
is significantly higher than older stellar populations.



\bsp	
\label{lastpage}
\end{document}